\documentclass[11pt]{article}
\usepackage{amsmath}
\usepackage{tensor}
\usepackage{amsfonts}
\usepackage[hmargin={27mm,27mm},vmargin={27mm,27mm}]{geometry}
\usepackage{hyperref}
\usepackage{verbatim}
\usepackage[shortlabels]{enumitem}
\usepackage{slashed}
\usepackage{braket}
\usepackage{caption}
\usepackage{cite}
\usepackage{nicefrac}
\usepackage{bbm}
\usepackage{url}
\usepackage{dsfont}

\numberwithin{equation}{section}

\def \ifempty#1{\def\temp{#1} \ifx\temp\empty }
\def \checkslot#1{ \ifempty{#1} \,\cdot\, \else #1 \fi }
\newcommand{\angl}[2]{\left\langle #1, #2 \right\rangle}
\newcommand{\trip}[3]{\left[\checkslot{#1},\checkslot{#2},\checkslot{#3}\right]}
\newcommand{\trips}[3]{\left[\checkslot{#1},\checkslot{#2};\checkslot{#3}\right]}
\newcommand{\bareps}{\bar{\epsilon}}

\begin{document}
 \centerline{\LARGE \bf {\sc  Non-Lorentzian RG flows and Supersymmetry} }

\centerline{\LARGE \bf {\sc  }} \vspace{2truecm} \thispagestyle{empty} \centerline{
    {\large {\bf {\sc Neil~Lambert }}}\footnote{E-mail address: \href{mailto:neil.lambert@kcl.ac.uk}{\tt neil.lambert@kcl.ac.uk}} and {\large {\bf {\sc Rishi Mouland }}}\footnote{E-mail address: \href{mailto:rishi.mouland@kcl.ac.uk}{\tt rishi.mouland@kcl.ac.uk}}  }

\vspace{1cm}
\centerline{ {\it Department of Mathematics}}
\centerline{{\it King's College London, }} 
\centerline{{\it  WC2R 2LS, UK}} 

\vspace{1.0truecm}

%%%%%%%%%%%%%%%%%
 
\thispagestyle{empty}

\centerline{\sc Abstract}
\vspace{0.4truecm}
\begin{center}
\begin{minipage}[c]{360pt}{
    }
We describe a general process where a non-Lorentzian rescaling of  a supersymmetric field theory  leads to a scale-invariant fixed point action without Lorentz invariance but where the supersymmetry is preserved or even enhanced. We apply this procedure to five-dimensional maximally supersymmetric super-Yang-Mills, leading to a field theory with 24 super(conformal) symmetries. We also apply this procedure to the BLG model with 32 super(conformal) symmetries and  ABJM models with 24 super(conformal) symmetries.

\end{minipage}
\end{center}

\newpage 
 
\renewcommand{\thefootnote}{\arabic{footnote}}
\setcounter{footnote}{0} 

\section{Introduction}

Supersymmetric field theories have long been of interest and are central to the dynamics of branes in String Theory and M-theory. Such theories are usually taken to be Lorentz invariant but non-Lorentzian supersymmetric theories can arise, typically in a system where one has fixed a frame of reference (such as in a DLCQ construction) or introduced a Lorentz-violating background. The brane dynamics are then often governed by a Lifshtiz or Galilean invariant field theory. Another class of examples are Carollian limits \cite{Duval:2014uoa}.  Similar models are also   of interest in condensed matter physics. Some examples of such theories are given in \cite{Leblanc:1992wu,Bandos:2008fr,Orlando:2009az,Xue:2010ih,Gomes:2014tua,Chapman:2015wha,Gallegos:2018jyg,Bagchi:2015qcw}. There is  also a considerable literature on the AdS duals of such theories   (for example see \cite{Balasubramanian:2008dm,Son:2008ye,Barbon:2008bg,Goldberger:2008vg,Herzog:2008wg,Adams:2008wt,Maldacena:2008wh,Donos:2010tu,Taylor:2015glc}).

Recently non-Lorentzian field theories with maximal supersymmetry were constructed by examining branes in M-theory in frames that have been infinitely boosted \cite{Lambert:2018lgt}.   The resulting dynamics localizes onto a moduli space of BPS states and can be further reduced to a quantum mechanical model.   In these theories the familiar Manton approximation for slow motion on a soliton moduli space becomes exact.  These appear to be a new class of  supersymmetric models with Lifshitz scaling symmetry. In particular, in distinction to traditional Lifshitz-like actions, they are at most quadratic in derivatives.

In this paper we consider a generic supersymmetric field theory and perform a non-Lorentzian scale transformation with a parameter $\eta$. For $\eta\ne 0$ the transformation is invertible and theory is equivalent to the original theory. However we construct the limiting theory at $\eta=0$ and show that supersymmetry can be preserved by removing divergent terms and replacing them with a  Lagrange multiplier constraint that sets them to zero.  This leads to a supersymmetric, scale-invariant fixed point theory without Lorentz symmetry (typically with just translations and spatial rotations). A similar process has been discussed for supergravity in \cite{Bergshoeff:2015uaa} although those authors pursue a different treatment of divergent terms.

This paper is organised as follows. In section two we present the general construction. In section three we apply this to the five-dimensional maximally supersymmetric Yang-Mills action and show that the resulting fixed point action has a Liftshitz scale symmetry and, in addition to the original 16 supersymmetries, 8 superconformal symmetries.  In section four we apply our techniques to the Chern-Simons-matter theories of M2-branes: at first the ${\cal N}=8$ BLG model and show that all 32 super and superconformal symmetries are preserved and then extend this to the ${\cal N}=6$ ABJM/ABJ models where all 24 super and superconformal symmetries are preserved. Section five contains our conclusions. We also include an appendix discussing how the particular scaling used in section three arises in the context of  M5-branes, leading to the proposal of \cite{Aharony:1997th,Aharony:1997an}.

\section{Scaling limits of supersymmetric actions}\label{sec: scaling limits}

\subsection{The basic case}

We start by outlining a general prescription by which we can take scaling limits of supersymmetric theories while retaining all of the initial supersymmetry. So  let $S$ be a supersymmetric action  and let $\{\Phi^\alpha\}$ be the set of all fields, both bosonic and fermionic, each taking values in some vector space $\mathcal{V}_\alpha$. For brevity and clarity, take the $\Phi^\alpha$ real, and let $(\,\,.\,\, ,\,\, .\,\,)$ be some real inner product on $\mathcal{V}=\oplus_\alpha\mathcal{V}_\alpha$. \\

Now suppose further that we introduce a continuous parameter $\eta$ and rescale all the fields and coordinates by some power of $\eta$; $\Phi^\alpha\to \eta^{\lambda_\alpha}\Phi^\alpha$, $x^\mu \to \eta^{\lambda_\mu}x^\mu$. The action may now be written
\begin{align}\label{Sexpand}
  S= \sum_{\lambda} \eta^\lambda S_\lambda\ ,
\end{align}
where each of the $S_\lambda$ is independent of $\eta$. In this paper we will restrict our attention to cases where, after a suitable choice of $\eta$ and scaling weights, the action just contains three terms:
\begin{align}
  S= \eta^{-1}S_{-1}+S_0 +\eta S_1\ ,
\end{align} 
although our results trivially extend to cases where there are an arbitrary number of 
terms with positive powers of $\eta$. For a fixed $\eta\ne 0$ nothing has really changed and the dynamics is equivalent to the undeformed case. Our aim here is to try to make sense of the theory in the limit that $\eta\to 0$. 

We also extract the $\eta$ dependance of the supersymmetry variations $\delta$ and we allow for the supersymmetry parameter  $\epsilon$ to also scale. In general these can be expanded in a similar expansion as (\ref{Sexpand}) but here we only consider cases where
\begin{align}
  \delta =  \eta^{-1} \delta_{-1} + \delta_0+\eta\delta_1\ .
\end{align}
Again our results also apply if there is an arbitrary number of terms with postive powers of $\eta$. 
Then, the fact that the action is supersymmetric, {\it i.e.} $\delta S=0$, can now be written as a tower of invariance equations for each $\lambda$:
\begin{align}
 \sum_{\lambda'} \delta_{\lambda'} S_{\lambda-\lambda'} = 0\ .
\end{align}
In particular we will have the invariance conditions:
\begin{align}\label{eq: InvCon}
  \delta_{-1} S_{-1} &= 0\nonumber\\
  \delta_{-1}S_0 + \delta_0 S_{-1} &=0\nonumber\\
  \delta_{-1}S_1 + \delta_0 S_0 + \delta_1 S_{-1} &= 0\ .
\end{align}
We introduce one final piece of notation. Just as the action has been split into a series in $\eta$, so can the equations of motion arising from the variation of each of the $\Phi^\alpha$. We let $E^{(\lambda)}_A$ denote the equation of motion for $\Phi^\alpha$ at level $\lambda$, {\it i.e.} under \textit{general} infinitesimal variations of the fields,
\begin{align}
  \delta S_{\lambda} = \int d^d x\,\, \left( \delta \Phi^\alpha , E^{(\lambda)}_\alpha\right)\ .
\end{align}
 
To analyse the limit  $\eta\to 0$  we consider the case where $S_{-1}$ is of the form
\begin{align}
  S_{-1} = \int d^d x\,\, \left[\frac{1}{2} \kappa^{ab} \left(\Omega_a, \Omega_b \right) \right]\ .
  \label{eq: simple S_{-1}}
\end{align}
Here  the $\Omega_a$ are generically composite fields made up of the $\Phi^\alpha$, and $\kappa^{ab}$ is symmetric and non-degenerate. The $\lambda=-2$ invariance condition implies that
\begin{align}\label{eq: S-2}
 \int d^d x\,\,\kappa^{ab}\left(\delta_{-1}\Phi^\alpha\frac{\partial \Omega_a}{\partial\Phi^\alpha},\Omega_b \right) =0\ .
\end{align}
To begin with we restrict our attention to the cases where $\Omega_a$ is purely bosonic and $\delta_{-1}\Phi^\alpha$ is only non-zero for fermionic fields. Thus (\ref{eq: S-2}) will be trivially satisfied as ${\partial \Omega_a}/{\partial\Phi^\alpha}=0$ for fermionic choices of $\alpha$. Our analysis can be extended to other cases (indeed we will study one such extension below). 

In the physically relevant cases where $S_{-1}$ is positive definite there is a straightforward interpretation of the theory in the limit $\eta\to 0$. We see that in this limit, we localise onto the classical minima of $S_{-1}$, while disregarding $S_\lambda$ for $\lambda > 0$. 
%Thus, if we are to provide a Lagrangian description of the limiting theory, it should be $S_0$ augmented by a Lagrange multiplier constraining the dynamics onto the classical minima of $S_{-1}$. 
As such it is important to make this constraint  as weak as possible by a suitable choice of boundary term in the definition of $S_{-1}$.  We can then  ignore such boundary terms as they do not contribute to the invariance of the action or on-shell dynamics (although they may still have a physical interpretation). The classical minima then simply correspond to $\Omega_a=0$. Thus  in the limit $\eta\to0$  we expect that the dynamics is captured by the simple action
\begin{align}
  \tilde S  = S_0+ \int d^d x \,\, \kappa^{ab}\left(  \Omega_a,G_b \right)\ ,
\end{align} 
where $G_a$ are new fields that impose the constraints $\Omega_a=0$. 
As we now show, under mild assumptions,  we suitably can modify the supersymmetry transformation $\delta \to \tilde \delta=\delta_0+\delta'$ so as to ensure that $\tilde \delta\tilde S=0$. \\

To establish this we observe that for such a form of $S_{-1}$, the $\lambda=-1$ invariance equation implies that
\begin{align}
  \delta_{-1}S_0 + \delta_0 S_{-1} = \int d^d x\,\, \left[ \left( \delta_{-1}\Phi^\alpha, E^{(0)}_\alpha \right) + \kappa^{ab}\left( \delta_0 \Omega_a, \Omega_b \right)   \right] = 0\ .
  \label{eq: -1 invariance equation}
\end{align}
We emphasise that this equation is satisfied purely algebraically and off-shell. If $\delta_{-1}\Phi^\alpha$ is only non-zero for fermionic fields and $\Omega_a$ is bosonic we see that $ E^{(0)}_\alpha$ and $\delta_0\Omega_a$ are fermionic. Therefore we can find functions $\Sigma^\alpha_a $ such that 
\begin{align} \delta_0\Omega_a &= - \Sigma_a^{\alpha} E^{(0)}_\alpha \ ,
\end{align}
and hence
\begin{align}
  \delta_{-1} \Phi^\alpha &= \kappa^{ab}{}^*\Sigma^\alpha_a  \Omega_b  \ ,\label{eq: Sigma}
\end{align}
where ${}^*\Sigma_b^{\alpha}$ is the adjoint map to $\Sigma_b^{\alpha}$ with respect to the $(\ ,\ )$ inner-product.
 With these relations in hand, we turn our attention to the $\lambda=0$ invariance equation,
\begin{align}
  \delta_0 S_0 + \delta_{-1} S_1 + \delta_1 S_{-1} = 0\ .
\end{align}
This means that $S_0$ is not invariant under the $\delta_0$ supersymmetry but rather 
\begin{align}
\delta_0S_0 &=  - \delta_{-1} S_1 - \delta_1 S_{-1} \nonumber\\
& = -\int d^d x\,\,  \kappa^{ab}\left({}^*\Sigma_b^{\alpha} \Omega_a,  E^{(1)}_\alpha\right) -\int d^d x\,\, \kappa^{ab}\left(\Omega_a, \delta_1  \Omega_b \right)\nonumber\\
& = -\int d^d x\,\,  \kappa^{ab}\left(\Omega_a,  \Sigma_b^{\alpha} E^{(1)}_\alpha\right) -\int d^d x\,\, \kappa^{ab}\left(\Omega_a, \delta_1 \Omega_b \right)\ ,
\end{align}

Our task now is to find the  corrected supersymmetry $\delta'$ to ensure that $\tilde \delta \tilde S=0$.  Since the   terms  in $\delta_0S_0$ are all proportional to $\Omega_a$ we can cancel them by taking
\begin{align}
\delta' G_a = \Sigma^\alpha_a E^{(1)}_\alpha + \delta_1\Omega_a\ .
\end{align}
This leaves us with
\begin{align}
\tilde\delta \tilde S & = \delta'S_0 + \int d^d x \,\, \kappa^{ab}\left( (\delta_0+\delta')\Omega_a,G_b \right)\nonumber\\
& =  \int d^d x \,\,(\delta' \Phi^\alpha, E_\alpha^{(0)})- \int d^d x \,\, \kappa^{ab}\left( \Sigma_a^\alpha E^{(0)}_\alpha,G_b \right) +   \int d^d x \,\, \kappa^{ab}\left(  \delta'\Omega_a,G_b \right)\ .
\end{align}
We can now set $\tilde\delta \tilde S=0$ by taking
\begin{align}
\delta' \Phi^\alpha &= {}^*\Sigma^\alpha_a G_b\kappa^{ab}\ ,
\end{align}
provided that we can also have $ \delta' \Omega_a  =0$. Namely we require that
\begin{align}
\delta' \Omega_a = \delta' \Phi^\alpha \frac{\partial \Omega_a}{\partial\Phi^\alpha} = {}^*\Sigma^\alpha_a G_b\kappa^{ab}\frac{\partial \Omega_a}{\partial\Phi^\alpha}=0\ .
\end{align}
Since $G_a$ is an independent field that doesn't appear in $\Omega_a$ or $ {}^*\Sigma^\alpha_a$ this implies
\begin{align}
{}^*\Sigma^\alpha_b  \frac{\partial \Omega_a}{\partial\Phi^\alpha}=0\ .\label{eq: condition}
\end{align}
When $\Omega_a$ is only made of bosonic fields  $\Sigma^\alpha_a$ is only non-zero for fermionic choices of the $\alpha$-index but ${\partial \Omega_a}/{\partial\Phi^\alpha}$ is only non-zero for bosonic choices of $\alpha$ and hence (\ref{eq: condition}) holds. 

Thus in summary, and in slightly more generality, we find that if the scaling leads to a divergent term in the action of the form (\ref{eq: simple S_{-1}}) then, assuming that we can construct the map $\Sigma^\alpha_a$ defined in (\ref{eq: Sigma}), we can construct a supersymmetric action $\tilde S$  where 
\begin{align}
 \tilde S  &= S_0+ \int d^d x \,\, \kappa^{ab}\left(  \Omega_a,G_b \right)\nonumber\\
 \tilde\delta \Phi^\alpha & = \delta \Phi^\alpha +{}^*\Sigma^\alpha_a G_b\kappa^{ab}\nonumber\\
 \tilde \delta G_a & = \Sigma^\alpha_a E^{(1)}_\alpha + \delta_1\Omega_a\ ,
\end{align}
provided that (\ref{eq: condition}) holds. In particular we have argued that this can always be done in the case that $S_{-1}$ is bosonic and $\delta_{-1}$ is fermionic. 
 
Lastly we point out that   we started off with a scaling  that was not a symmetry of the original action $S$, otherwise we would have $S_0=S$, $\delta_0=\delta $ and hence $\tilde S=S$, $\tilde \delta=\delta$, meaning that the whole process was trivial. However this scaling is a symmetry of $\tilde S$. In particular $S_0$ is scale invariant by construction and, in order to have appeared in $S_{-1}$, $\Omega_a$ must  scale  as $\Omega_a\to \eta^{-(\lambda_0+\lambda_1\dots+\lambda_{d-1}+1)/2}\Omega_a$. Thus if we let $G_a\to \eta^{-(\lambda_0+\lambda_1\ldots+\lambda_{d-1}-1)/2}G_a $ then $\tilde S$ will be  invariant under the scale transformation. 

In other words, by introducing a scaling we have produced an inhomogeneous RG flow resulting in a fixed point theory in the limit $\eta\to 0$, which is invariant under the scaling.

\subsection{A more general prescription}\label{subsec: more general prescription}

The crucial quality of the initial theory that allowed for this procedure to work was the quite simple form (\ref{eq: simple S_{-1}}) of $S_{-1}$. A related condition is that  $\delta_{-1}\Phi^\alpha\neq 0$ only for  fermionic $\Phi^\alpha$. We can adapt this prescription for a more general situation as we now discuss, although the details are better left to the explicit examples below.

We saw that the scaling led to an RG flow with $\tilde S$ the fixed point action. The flow is rather trivial in that all the fields simply obey their naive scaling behaviour. However one could allow for field redefinitions and mixings along the flow. Thus we could consider making the field redefinition 
\begin{align}
\Phi^\alpha \to \Phi^\alpha - \eta^{-1}\chi^\alpha\ ,
\end{align}
in the rescaled theory, 
where $\chi^\alpha$ is some function of the fields.\footnote{Again one could allow for more general powers of $\eta$ but we leave that to future work. Here we assume the same  expansion in terms of $\eta^{-1}$, $\eta^0$ and $\eta$ that we considered above.} This has two key effects. Firstly it leads to a shift in $\delta_{-1}\Phi^\alpha$:
\begin{align}
\delta_{-1}\Phi^\alpha \to \delta_{-1}\Phi^\alpha - \delta_0\chi^\alpha\ ,
\end{align} 
and secondly it will affect the form of $S_{-1}$:
\begin{align}
S_{-1} \to S_{-1}- \int d^d x \,\, \left(\chi^\alpha,E^{(0)}_\alpha\right) +   \frac 12 \int d^d x \,\, \left(\chi^\alpha\chi^\beta,\frac{\partial E^{(1)}_\alpha}{\partial\Phi^\beta}\right)\ .
\end{align} 
as well as  shifting other terms around. In general this will introduce  $S_{-2}$ and $\delta_{-2}$ and even more divergent terms. These would in turn lead to additional terms in the invariance conditions (\ref{eq: InvCon}) and invalidate the previous discussion. However we will see   below that there are  cases where a suitable choice of field redefinition   maps the theory back to the situation studied   above where $S_{-1}$ takes the form of (\ref{eq: simple S_{-1}})  and $\delta_{-1}\Phi^\alpha\ne 0$ only for fermions, without introducing higher divergences.

This provides a procedure for cases the where $\delta_{-1}\Phi^\alpha\ne 0$ for bosons. Namely one first looks for a field definition such that $\delta_{-1}\Phi^\alpha= 0$ for the new bosonic fields and then determines the resulting form of $S_{-1}$. If it is of the form (\ref{eq: simple S_{-1}}), with no further divergences, then one can proceed as in the previous discussion.

\section{Yang-Mills}\label{sec: M5s}

The simplest application of the above construction starts with five-dimensional maximally supersymmetric Yang-Mills:
\begin{align}
S = \frac{1}{g^2} \text{tr}\int d^5 x\, &\left(  - \frac{1}{4} F_{\mu\nu} F^{\mu\nu} -\frac{1}{2} D_{\mu }X^I D^{\mu } X^I + \frac{1}{4}[X^I,X^J][X^I,X^J]\right.  \nonumber\\ 
&\left. +\frac{i}{2}\bar\Psi\Gamma^\mu D_\mu\Psi -\frac{1}{2}\bar\Psi\Gamma_5 \Gamma^I[X^I,\Psi]\right)\ .
\end{align}
In this section we take $\mu,\nu=0,1,2,3,4$, $I,J=6,7,8,9,10$ and the $\Gamma$-matrices are a real $32\times 32$ representation of $Spin(1,10)$. Additionally, the fermions satisfy $\Gamma_{012345}\Psi=-\Psi$. This action has 
 the supersymmetry
\begin{align}
\delta X^I & = i\bar\xi \Gamma^I\Psi\nonumber\\
\delta A_\mu &= i\bar\xi\Gamma_\mu\Gamma_5\Psi\nonumber\\
\delta \Psi & = \frac12 \Gamma^{\mu\nu}\Gamma_5F_{\mu\nu}\xi + \Gamma^\mu\Gamma^ID_\mu X^I \xi  - \frac{i}{2}\Gamma^{IJ}\Gamma_5[X^I,X^J]\xi\ .
\label{eq: YM SUSYs}
\end{align}
with $\Gamma_{012345}\xi=\xi$. Next we want to consider a rescaling of the theory. In particular we take
\begin{align}\label{eq: Lifshitz}
x^0&\to \eta^{-1}  x^0\nonumber\\
x^i&\to \eta^{-1/2} x^i\nonumber\\
X^I&\to   \eta X^I\nonumber\\
\Psi_+ &\to  \eta \Psi_+\nonumber\\
\Psi_- &\to  \eta^{3/2} \Psi_-\nonumber\\ 
\xi_+ &\to \eta^{-1/2}\xi_+\nonumber\\
\xi_-&\to  \eta^0 \xi_-\ ,
\end{align}
where $i,j=1,2,3,4$ and we have introduced
\begin{align}
  \Gamma_\pm = \frac{1}{\sqrt{2}}\left( \Gamma_0 \pm \Gamma_5 \right)\ ,
\end{align}
and the corresponding projections 
\begin{align}
\Psi_{\pm} = \frac12(1\pm \Gamma_{05})\Psi\qquad  \xi_{\pm} = \frac12(1\pm \Gamma_{05})\xi\ .
\end{align} 
One can motivate this particular scaling by considering the action on multiple M5-branes in an infinitely boosted frame, as explained in the appendix.

Following the discussion above this scaling leads to 
\begin{align}
  S = \eta^{-1} S_{-1} + S_0 + \eta^1 S_1\ ,
\end{align}
where  (we also further rescale  all the positive chirality spinors by a factor of $2^{1/4}$ and negative chirality ones by $2^{-1/4}$ so as to agree with \cite{Lambert:2018lgt})\begin{align}
  S_{-1} &= \frac{1}{g^2} \text{tr}\int d^4 x\, dx^0 \left( -\frac{1}{4} F_{ij} F_{ij} \right)  \nonumber\\ 
  S_0 &=  \frac{1}{g^2} \text{tr}\int d^4 x\, dx^0 \Bigg( \frac{1}{2} F_{0i} F_{0i} - \frac{1}{2} \left( D_i X^I \right) \left( D_i X^I \right) - \frac{i}{2} \bar{\Psi} \Gamma_- D_0 \Psi \nonumber\\
  &\hspace{36mm}  +\frac{i}{2} \bar{\Psi} \Gamma_i D_i \Psi + \frac{1}{2} \bar{\Psi} \Gamma_- \Gamma^I [X^I,\Psi] \Bigg)\nonumber\\
  S_{1} &= \frac{1}{g^2} \text{tr}\int d^4 x\, dx^0 \Bigg( \frac{1}{2} \left( D_0 X^I \right) \left( D_0 X^I \right) - \frac{i}{4}\bar{\Psi} \Gamma_+ D_0 \Psi \nonumber\\
  &\hspace{36mm} -\frac{1}{4} \bar{\Psi} \Gamma_+ \Gamma^I [X^I, \Psi] + \frac{1}{4} [X^I, X^J][X^I, X^J] \Bigg)\ .
  \label{eq: YM scaled}
\end{align}
The  supersymmetries take the form
\begin{align}
  \delta = \eta^{-1} \delta_{-1} + \delta_0 + \eta \delta_{1}\ ,
\end{align}
with
\begin{align}
  \delta_{-1} \Psi &= -\tfrac{1}{2} F_{ij} \Gamma_{ij} \Gamma_- \xi  \\[1em]
  \delta_0 X^I &= i \bar{\xi} \Gamma^I \Psi \nonumber \\
  \delta_0 A_0 &= i \bar{\xi} \Gamma_{-+} \Psi \nonumber \\
  \delta_0 A_i &= -i\bar{\xi} \Gamma_i \Gamma_- \Psi %\quad  \implies \quad \delta_0 F_{ij} = -2i \bar{\epsilon} \Gamma_- \Gamma_{[i} D_{j]} \Psi \nonumber
   \\
  \delta_0 \Psi &= F_{0i} \Gamma_i \Gamma_{-+}\xi + \tfrac{1}{4} F_{ij} \Gamma_{ij} \Gamma_+ \xi + \left( D_0 X^I \right) \Gamma^I \Gamma_- \xi \nonumber \\
  &\hspace{8mm}+ \left( D_i X^I \right) \Gamma_i \Gamma^I \xi + \tfrac{i}{2} [X^I, X^J] \Gamma^{IJ} \Gamma_- \xi \nonumber \\[1em]
  \delta_1 A_i &= \tfrac{i}{2} \bar{\xi} \Gamma_i \Gamma_+ \Psi% \quad \implies \quad \delta_1 F_{ij} = i\bar{\epsilon} \Gamma_+ \Gamma_{[i} D_{j]} \Psi 
  \nonumber \\
  \delta_1 \Psi &= \tfrac{1}{2} \left( D_0 X^I \right) \Gamma^I \Gamma_+ \xi - \tfrac{i}{4} [X^I, X^J] \Gamma^{IJ} \Gamma_+ \xi\ .
  \label{eq: YM scaled SUSYs}
\end{align}

We want to make the constraint $S_{-1}=0$ as weak as possible so we shift $S_{-1}$ by a topological piece proportional to $\varepsilon^{ijkl}{\rm tr}(F_{ij}   F_{kl})$   to obtain
\begin{align}
 S_{-1} &= -\frac{1}{2g^2} \text{tr}\int d^4 x\, dx^0 \left( F^-_{ij} F^-_{ij} \right)\ ,
\end{align}
where $F^-_{ij}=\frac12 F_{ij}-\frac{1}{4}\varepsilon_{ijkl}F_{kl}$.
In terms of the notation above we have $a\to[ij]$, $\kappa_{ab}\to\kappa_{ij,kl}=-\delta_{ik}\delta_{jl}$ and $\Omega_{ij} = \frac{1}{g}F_{ij}^-$. We then find
\begin{align}
\delta_0\Omega_{ij} & =-\frac{i}{2g} \bar\xi \Gamma_-\Gamma_{ij}\Gamma_kD_k\Psi = -\frac{g}{2} \bar\xi \Gamma_-\Gamma_{ij} E^{(0)}_{\Psi} \nonumber\\
\delta_{-1} \Psi &= -\frac{1}{2} F_{ij}^- \Gamma_{ij} \Gamma_- \xi\ = -\frac{g}{2} \Omega_{ij} \Gamma_{ij} \Gamma_- \xi\ ,
\end{align}
Hence, if we take $\Sigma_{ij}^\Psi  =  \tfrac{g}{2} \bar\xi  \Gamma_- \Gamma_{ij}$, with $\Sigma^\alpha_{ij}=0$ for all $\alpha\neq\Psi$, we indeed have
\begin{align}
 \delta_0 \Omega_{ij} &= -\Sigma_{ij}^\Psi E_\Psi^{(0)}\nonumber\\
 \delta_{-1} \Psi &= \kappa^{ij,kl} \, {}^*\Sigma_{ij}^\Psi \Omega_{kl} = -{}^*\Sigma_{ij}^\Psi \Omega_{ij} \ ,
\end{align}
where here ${}^*$ is simply the Dirac conjugate.
So, we are safe to proceed with the procedure. The end result is that, in the limit $\eta\to 0$, the theory is described by the following action 
\begin{align}
  S = \frac{1}{g^2} \text{tr}\int d^4 x\, dx^0 \Bigg( \frac{1}{2} F_{0i} F_{0i} + \frac{1}{2} F_{ij} G_{ij} - \frac{1}{2} \left( D_i X^I \right) \left( D_i X^I \right) \nonumber\\
  - \frac{i}{2} \bar{\Psi} \Gamma_- D_0 \Psi  +\frac{i}{2} \bar{\Psi} \Gamma_i D_i \Psi + \frac{1}{2} \bar{\Psi} \Gamma_- \Gamma^I [X^I,\Psi] \Bigg)\ ,
  \label{eq: final M5 action}
\end{align}
where the new Lagrange multiplier is an anti-self-dual spatial 2-form $G_{ij}$. The supersymmetry variations are
\begin{align}
  \delta X^I &= i \bar{\xi} \Gamma^I \Psi \nonumber \\
  \delta A_0 &= i \bar{\xi} \Gamma_{-+} \Psi \nonumber \\
  \delta A_i &= -i\bar{\xi} \Gamma_i \Gamma_- \Psi \quad \nonumber \\
  \delta \Psi &= F_{0i} \Gamma_i \Gamma_{-+}\xi + \tfrac{1}{4} F_{ij} \Gamma_{ij} \Gamma_+ \xi + \left( D_0 X^I \right) \Gamma^I \Gamma_- \xi \nonumber \\
  &\hspace{8mm}+ \left( D_i X^I \right) \Gamma_i \Gamma^I \xi + \tfrac{i}{2} [X^I, X^J] \Gamma^{IJ} \Gamma_- \xi + \tfrac{1}{4} G_{ij} \Gamma_{ij} \Gamma_- \xi \nonumber\\
  \delta G_{ij} &= \tfrac{i}{2}\bar{\xi}\Gamma_+ \Gamma_k \Gamma_{ij} D_k \Psi + \tfrac{i}{2}\bar{\xi}\Gamma_- \Gamma_{ij} \Gamma_+ D_0 \Psi + \tfrac{1}{2} \bar{\xi}\Gamma_- \Gamma_{ij} \Gamma_+ \Gamma^I [X^I, \Psi]\ .
\end{align}
This reproduces the theory first obtained   in \cite{Lambert:2018lgt} (but the equations of motion where found and analysed in \cite{Lambert:2010wm}). Upon integrating out $G_{ij}$  we are stricted to the space of self-dual gauge fields on ${\mathbb R}^4$ and the theory reduces to motion on instanton moduli space, see \cite{Lambert:2011gb}. 

It is easy to check that (\ref{eq: final M5 action}) has the Liftshitz scaling symmetry
(\ref{eq: Lifshitz}) provided that $G_{ij}\to \eta^2 G_{ij}$. However one can also see that it has an additional superconformal symmetry  which does not have an analogue in the original theory:
\begin{align}
  \delta X^I &= i \bar{\epsilon} \Gamma^I \Psi \nonumber \\
  \delta A_0 &= i \bar{\epsilon} \Gamma_{-+} \Psi \nonumber \\
  \delta A_i &= -i\bar{\epsilon} \Gamma_i \Gamma_- \Psi \quad \nonumber \\
  \delta \Psi &= F_{0i} \Gamma_i \Gamma_{-+}\epsilon  + \tfrac{1}{4} F_{ij} \Gamma_{ij} \Gamma_+ \epsilon  + \left( D_0 X^I \right) \Gamma^I \Gamma_- \epsilon + \left( D_i X^I \right) \Gamma_i \Gamma^I \epsilon  \nonumber \\
  &\hspace{8mm}+ \tfrac{i}{2} [X^I, X^J] \Gamma^{IJ} \Gamma_- \epsilon + \tfrac{1}{4} G_{ij} \Gamma_{ij} \Gamma_- \epsilon -4X^I\Gamma^I\zeta_-\nonumber\\
  \delta G_{ij} &= \tfrac{i}{2}\bar{\epsilon}\Gamma_+ \Gamma_k \Gamma_{ij} D_k \Psi + \tfrac{i}{2}\bar{\epsilon}\Gamma_- \Gamma_{ij} \Gamma_+ D_0 \Psi + \tfrac{1}{2} \bar{\epsilon}\Gamma_- \Gamma_{ij} \Gamma_+ \Gamma^I [X^I, \Psi] + 3i\bar\zeta_-\Gamma_{+}\Gamma_{ij}\Psi\ ,
\end{align}
where now  $\epsilon=\xi +  x^0\Gamma_+\zeta_- +x^i\Gamma_i\zeta_-$. Thus there are 16 supersymmetries parameterized by a constant $\xi$ and an additional 8 superconformal supersymmetries parameterized by a constant $\zeta_-$. These are consistent with the 32 super(conformal) symmetries of the M5-brane theory reduced on a null direction $x^-$ with the restriction that all fields and supersymmetries are independent of $x^-$ (which explains why there is no $\zeta_+$ superconformal symmetry as the resulting $\xi$ would be linear in $x^-$).

\subsection{A 1-parameter family of alternatives}

We could ask what would happen had we not shifted $S_{-1}$ by a topological piece. Indeed, there is a 1-parameter family of ways in which we can write $S_{-1}$, each related by a shift by some multiple of $\varepsilon^{ijkl}{\rm tr}(F_{ij} F_{kl})$. It's easily seen that the form we chose is the unique choice such that the integrand (\ref{eq: -1 invariance equation}) vanishes identically. Otherwise, it is equal to a boundary term, thanks to the Bianchi identity for $F_{ij}$.\\

What this  boils down to is the fact that, for other choices of the boundary term,  we must also relax the condition that $G_{ij}$ be anti-self-dual and allow it to be a general spatial 2-form. One finds that  $\delta G_{ij}$  is then shifted by
\begin{align}
  \delta G_{ij} \to \delta G_{ij} + i\alpha  \bar{\epsilon}\Gamma_+  \Gamma_{ijk} D_k \Psi = \delta G_{ij} + \tfrac{i\alpha}{2} ( \underbrace{\bar{\epsilon}\Gamma_+  \Gamma_{ij}\Gamma_{k} D_k \Psi}_{\text{self-dual}} + \underbrace{\bar{\epsilon}\Gamma_+  \Gamma_{k}\Gamma_{ij} D_k \Psi}_{\text{anti-self-dual}} )\ ,
\end{align}
for some $\alpha\in\mathbb{R}$. For any $\alpha\neq 0$, $\delta G_{ij}$ has both non-zero self-dual and anti-self-dual parts, and we are forced to regard $G_{ij}$ as a generic 2-form. Upon integrating out such a $G_{ij}$, we have the flat connection condition $F_{ij}=0$.

Thus, the theory described by the action (\ref{eq: final M5 action}) can be seen as a special case, where $\alpha=0$, $\delta G^+_{ij}=0$, and we are safely able to regard $G_{ij}$ as anti-self-dual. As we've already mentioned, integrating out $G_{ij}$ puts us onto instanton moduli space, with instanton number $k\ge 0$. Of course in the trivial sector ($k=0$), the condition is simply $F_{ij}=0$. Thus, we haven't lost anything by going to this special case, $\alpha=0$. Notably, however, we see that only the instanton sectors (as opposed to anti-instanton) are accessible. Of course the reverse would be the case if we'd instead chosen $x^+$ as our null-compactified direction.

\section{Chern-Simons-Matter}\label{sec: M2s}

We now turn our attention to similar limits of worldvolume theories for multiple M2-branes. In \cite{Kucharski:2017jwv}, a non-Lorentzian variant of the ${\cal N}=8$ BLG theory was constructed  from the $(2,0)$ system of \cite{Lambert:2010wm,Lambert:2016xbs}. It was further explained that this system is U-dual to the DLCQ description of M5-branes. As we have just seen, the latter can be obtained via a scaling limit of five-dimensional maximally supersymmetric Yang-Mills. Thus   we expect that a similar scaling limit of the ${\cal N}=8$  theory will obtain the theory of \cite{Kucharski:2017jwv}.  

We first look at the ${\cal N}=8$  theory, where it turns out there is essentially a unique way in which we must scale the fields and coordinates in order to arrive at the theory of \cite{Lambert:2018lgt}. In section \ref{subsec: ABJM}, we apply the same scaling to the 
ABJM/ABJ action, and thus derive the associated  non-Lorentzian fixed-point theory, which nonetheless still has manifest $\mathcal{N}=6$ supersymmetry.

\subsection{${\cal N}=8$}\label{subsec: BLG}

Our starting point is the regular ${\cal N}=8$  theory \cite{Bagger:2007jr,Gustavsson:2007vu}. The dynamical fields take values in a three-algebra $\mathcal{V}$ with invariant inner product $\langle \, \cdot \, , \, \cdot \, \rangle$ and totally anti-symmetric product
\begin{align}
  [\,\cdot\, , \,\cdot\, , \,\cdot\, ]: \mathcal{V} \otimes \mathcal{V} \otimes \mathcal{V} \to \mathcal{V}\ ,
\end{align}
which acts on itself as a derivation,
\begin{align}
  [U,V,[X,Y,Z]] = [[U,V,X],Y,Z] + [X,[U,V,Y],Z] + [X,Y,[U,V,Z]]\ .
\end{align}
Additionally, the three-algebra generates a Lie-algebra $\mathcal{G}$ by the analogue of the adjoint map, $X\to \varphi_{U,V}(X) = [U,V,X]$ for any $U,V\in \mathcal{V}$. This naturally induces an invariant inner product $( \, \cdot \, , \, \cdot \, )$ on $\mathcal{G}$, which satisfies
\begin{align}
  (T,\varphi_{U,V}) = \langle T(U), V \rangle \ ,
\end{align}
for any $T\in\mathcal{G}$ and $U,V\in \mathcal{V}$.

The action is
\begin{align}
  S = \int d^3 x\,\, \bigg( -\frac{1}{2} \angl{ D_\mu X^{I}}{ D^\mu X^{I} } - \frac{1}{12} \angl{\trip{X^{I}}{X^{J}}{X^{K}}}{\trip{X^{I}}{X^{J}}{X^{K}}} +\frac{i}{2} \angl{\bar{\Psi}}{\Gamma^\mu D_\mu \Psi}\nonumber\\
   + \frac{i}{4} \angl{\bar{\Psi}}{\Gamma^{{I}{J}}\trip{X^{I}}{X^{J}}{\Psi}} + \frac{1}{2} \epsilon^{\mu\nu\lambda}\left( \left( A_m, \partial_n A_p \right) - \frac{2}{3}\left( A_\mu , A_\nu A_\lambda \right) \right)\bigg)\ ,
   \label{eq: plain BLG action}
\end{align}
where in this section $\mu,\nu,\lambda=0,1,2$, and we use the convention $\epsilon^{012}=-1$. The scalars $X^{I}$, ${I}=3,4,\dots,10$ and 32-component real spinors $\Psi$ take values in $\mathcal{V}$, while the gauge field $A_\mu$ takes values in $\mathcal{G}$. We take the field strength to be $F_{\mu\nu}=-[D_\mu,D_\nu] = \partial_\mu A_\nu - \partial_\nu A_\mu - [A_\mu,A_\nu]$. The spinors additionally satisfy $\Gamma_{012}\Psi = -\Psi$. We have
\begin{align}
  D_\mu X^{I} = \partial_\mu X^{I} - A_\mu \left( X^{I} \right)\ ,
\end{align}
and similarly for $\Psi$. The theory possesses 16 supercharges corresponding to rigid supersymmetry, and an additional 16 corresponding to superconformal symmetry. In particular, we have $\delta S=0$, with
\begin{align}
  \delta X^{I} &= i \bar{\epsilon} \Gamma^{I} \Psi\nonumber\\
  \delta \Psi &= \left( D_\mu X^{I} \right) \Gamma^\mu \Gamma^{I} \epsilon - \tfrac{1}{6} \trip{X^{I}}{X^{J}}{X^{K}} \Gamma^{{I}{J}{K}} \epsilon  -\tfrac{1}{3}X^I \Gamma^I \Gamma^\mu \partial_\mu\epsilon  \nonumber\\
  \delta A_\mu ( \,\cdot\, ) &= i\bar{\epsilon} \Gamma_\mu \Gamma^{I} \trip{X^{I}}{\Psi}{\,\cdot\,}\ ,
\end{align}
where $\epsilon$ takes the form $\epsilon=\xi + x^\mu \Gamma_\mu \zeta$. We additionally have $\Gamma_{012}\xi  = \xi$ and $\Gamma_{012}\zeta  = \zeta$, and so $\Gamma_{012}\epsilon = \epsilon$. 

Motivated by the theory obtained in \cite{Lambert:2018lgt}, we first introduce complex coordinate on the worldvolume $z=x^1+ix^2$. We also combine two of the off-brane scalars into a complex pair, $Z=X^3+iX^4$, while labelling the rest $X^A$ with $A=5,6,\dots,10$. Further, we split spinors into definite chiralities under the $\Gamma_{034} = -2i \Gamma_0 \Gamma_{Z\bar{Z}}$, so that {\it e.g.} $\Psi_\pm := \frac{1}{2}\left( 1\pm \Gamma_{034} \right)\Psi$.

Given this form of $S$ and $\delta$, we consider the scaling
\begin{align}
  x^0 &\to \eta^{-1} x^0 & \xi_+ &\to \eta^{-1/2} \xi_+ \nonumber \\
  z,\bar{z} &\to \eta^{- 1/2} z,\bar{z} & \xi_- &\to \xi_-\nonumber\\
  Z,\bar{Z} &\to Z,\bar{Z} & \zeta_+ &\to \eta^{1/2} \zeta_+\nonumber\\
  X^A &\to \eta^{1/2} X^A & \zeta_- &\to \eta  \zeta_-\nonumber\\
  \Psi_+ &\to \eta^{1/2} \Psi_+\nonumber\\
  \Psi_- &\to \eta \Psi_-\ .
  \label{eq: BLG scaling}
\end{align}
The scaling of $\xi$ and $\zeta$ imply that it terms of $\epsilon_\pm$  we have
\begin{align}
  \epsilon_+ &\to \eta^{-1/2} \left( \left( 1-z\partial -\bar{z}\bar{\partial} \right)\epsilon_+ + \eta  \left( z\partial +\bar{z}\bar{\partial} \right)\epsilon_+\right) \nonumber\\
  \epsilon_- &\to \epsilon_-\ .
\end{align}
After this scaling, the action is of the form $S=\eta^{-1} S_{-1} + S_0 + \eta S_1$, with
\begin{align}
  S_{-1} &= \int d^3 x\, \bigg( -\angl{DZ}{\bar{D}\bar{Z}} - \angl{D\bar{Z}}{\bar{D}Z} + \frac{1}{8} \angl{\trip{X^A}{Z}{\bar{Z}}}{\trip{X^A}{Z}{\bar{Z}}}
  + \frac{1}{4} \angl{\bar{\Psi}_+}{\Gamma_0\trip{Z}{\bar{Z}}{\Psi_+}}\bigg)  \\
  S_0 &= \int d^3 x\, \bigg( \frac{1}{2} \angl{D_0 Z}{D_0 \bar{Z}} - 2 \angl{DX^A}{\bar{D}X^A} - \frac{1}{4} \angl{\trip{X^A}{X^B}{Z}}{\trip{X^A}{X^B}{\bar{Z}}} \\
  &\hspace{10mm}  -\frac{i}{2} \angl{\bar{\Psi}_+}{\Gamma_0 D_0 \Psi_+} + 2i \angl{\bar{\Psi}_+}{\left( \Gamma_z \bar{D} + \Gamma_{\bar{z}}D\right)\Psi_-} - \frac{1}{4}\angl{\bar{\Psi}_-}{\Gamma_0 \trip{Z}{\bar{Z}}{\Psi_-}}\nonumber\\
  &\hspace{10mm} +i\angl{\bar{\Psi}_+}{\Gamma^A\Gamma_Z\trip{X^A}{Z}{\Psi_-}} + i\angl{\bar{\Psi}_+}{\Gamma^A\Gamma_{\bar{Z}}\trip{X^A}{\bar{Z}}{\Psi_-}} + \frac{i}{4} \angl{\bar{\Psi}_+}{\Gamma^{AB}\trip{X^A}{X^B}{\Psi_+}} \nonumber\\
  &\hspace{10mm} + i \big( \left( A_0, F_{z\bar{z}} \right) + \left( A_{\bar{z}}, F_{0z} \right) + \left( A_z, F_{\bar{z}0} \right) + \left( A_0, \left[ A_z, A_{\bar{z}} \right] \right) \big) \bigg) \nonumber\\
  S_{1} &= \int d^3 x\, \bigg( \frac{1}{2} \angl{D_0 X^A}{D_0 X^A} - \frac{1}{12} \angl{\trip{X^A}{X^B}{X^C}}{\trip{X^A}{X^B}{X^C}}\nonumber\\
  &\hspace{10mm} - \frac{i}{2}\angl{\bar{\Psi}_-}{\Gamma_0 D_0 \Psi_-} + \frac{i}{4}\angl{\Bar{\Psi}_-}{\Gamma^{AB}\trip{X^A}{X^B}{\Psi_-}} \bigg)\ .
  \label{eq: BLG pre-redef}
\end{align}
This action is invariant under $\delta = \eta^{-1} \delta_{-1} + \delta_0 + \eta \delta_1$, with
\newcommand{\oneminz}{\left( 1-z\partial \right)}
\newcommand{\oneminzbar}{\left( 1-\bar{z}\bar{\partial} \right)}
\newcommand{\oneminboth}{\left( 1-z\partial -\bar{z}\bar{\partial} \right)}
\begin{align}
  \delta_{-1} \Psi_- &= 2 \left( \bar{D}Z \right) \Gamma_z \Gamma_Z \oneminzbar\epsilon_+ + 2\left( D\bar{Z} \right)\Gamma_{\bar{z}}\Gamma_{\bar{Z}}\oneminz\epsilon_+  \\
  &\qquad +\tfrac{i}{2} \trip{X^A}{Z}{\bar{Z}}\Gamma^A \Gamma_0\oneminboth\epsilon_+\nonumber\\
  \delta_{-1} A_0 (\,\cdot\,) &= -\bar{\epsilon}_+ \Gamma_Z \trip{Z}{\Psi_+}{} + \bar{\epsilon}_+ \Gamma_{\bar{Z}} \trip{\bar{Z}}{\Psi_+}{}\nonumber\\[1em]
  \delta_0 Z &= 2i\left( \oneminz\bar{\epsilon}_+ \right) \Gamma_{\bar{Z}}\Psi_+  \\
  \delta_0 \bar{Z} &= 2i\left( \oneminzbar\bar{\epsilon}_+ \right) \Gamma_{Z}\Psi_+  \nonumber\\
  \delta_0 X^A &= i\left(\oneminboth \bar{\epsilon}_+ \right) \Gamma^A \Psi_- + i\bar{\epsilon}_- \Gamma^A \Psi_+  \nonumber\\
  \delta_0 \Psi_+ &= - \left( D_0 Z \right) \Gamma_0 \Gamma_Z \oneminzbar\epsilon_+ - \left( D_0 \bar{Z} \right) \Gamma_0 \Gamma_{\bar{Z}} \oneminz\epsilon_+ + 2 \left( DZ \right) \Gamma_{\bar{z}}\Gamma_Z \epsilon_-  \nonumber\\
  &\qquad + 2 \left( \bar{D}\bar{Z} \right) \Gamma_{z}\Gamma_{\bar{Z}} \epsilon_- + 2 \left( DX^A \right) \Gamma_{\bar{z}} \Gamma^A \oneminz\epsilon_+ + 2 \left( \bar{D} X^A \right) \Gamma_{z} \Gamma^A \oneminzbar\epsilon_+  \nonumber\\
  &\qquad -\tfrac{1}{2} \trip{X^A}{X^B}{Z}\Gamma^{AB} \Gamma_Z \oneminzbar\epsilon_+ -\tfrac{1}{2} \trip{X^A}{X^B}{\bar{Z}}\Gamma^{AB} \Gamma_{\bar{Z}} \oneminz\epsilon_+ \nonumber\\
  &\qquad - \tfrac{i}{2}\trip{X^A}{Z}{\bar{Z}}\Gamma^A \Gamma_0\epsilon_- + \tfrac{2}{3} Z\Gamma_{\bar{z}} \Gamma_Z \partial\epsilon_- + \tfrac{2}{3} \bar{Z}\Gamma_{z} \Gamma_{\bar{Z}} \bar{\partial}\epsilon_- \nonumber\\
  &\qquad + \tfrac{1}{3}Z\Gamma_Z \Gamma_0 \partial_0 \epsilon_+ + \tfrac{1}{3} \bar{Z}\Gamma_{\bar{Z}}\Gamma_0 \partial_- \epsilon_+ \nonumber\\
  \delta_0 \Psi_- &= - \left( D_0 X^A \right) \Gamma_0 \Gamma^A \oneminboth\epsilon_+ + 2 \left( D X^A \right) \Gamma_{\bar{z}}\Gamma^A \epsilon_- + 2 \left( \bar{D} X^A \right) \Gamma_{z}\Gamma^A \epsilon_- \nonumber\\
  &\qquad - \left( D_0 Z \right) \Gamma_0\Gamma_Z \epsilon_- -\left( D_0 \bar{Z} \right) \Gamma_0\Gamma_{\bar{Z}} \epsilon_- -\tfrac{1}{6}\trip{X^A}{X^B}{X^C}\Gamma^{ABC}\oneminboth\epsilon_+ \nonumber\\
  &\qquad -\tfrac{1}{2}\trip{X^A}{X^B}{Z}\Gamma^{AB}\Gamma_Z\epsilon_- - \tfrac{1}{2}\trip{X^A}{X^B}{\bar{Z}}\Gamma^{AB}\Gamma_{\bar{Z}}\epsilon_- + 2\bar{z}\left( \bar{D}Z \right) \Gamma_z \Gamma_Z \bar{\partial}\epsilon_+ \nonumber\\
  &\qquad + 2z\left( D\bar{Z} \right) \Gamma_{\bar{z}} \Gamma_{\bar{Z}} \partial\epsilon_+ + \tfrac{i}{2} \trip{X^A}{Z}{\bar{Z}} \Gamma^A \Gamma_0 \left( z\partial + \bar{z}\bar{\partial} \right) \epsilon_+ + \tfrac{2}{3} Z\Gamma_z\Gamma_Z\bar{\partial}\epsilon_+ \nonumber\\
  &\qquad + \tfrac{2}{3} \bar{Z}\Gamma_{\bar{z}}\Gamma_{\bar{Z}}\partial\epsilon_+ + \tfrac{1}{3} Z\Gamma_Z \Gamma_0 \partial_0 \epsilon_- + \tfrac{1}{3} \bar{Z} \Gamma_{\bar{Z}} \Gamma_0 \partial_0 \epsilon_- \nonumber\\
  &\qquad - \tfrac{2}{3} X^A \Gamma^A \Gamma_{\bar{z}} \partial \epsilon_- - \tfrac{2}{3} X^A \Gamma^A \Gamma_{z} \bar{\partial} \epsilon_- +\tfrac{1}{3} X^A \Gamma^A \Gamma_0 \partial_0 \epsilon_+ \nonumber\\
  \delta_0 A_0(\,\cdot\,) &= \bar{\epsilon}_- \Gamma_Z \trip{Z}{\Psi_-}{} - \bar{\epsilon}_- \Gamma_{\bar{Z}} \trip{\bar{Z}}{\Psi_-}{} \nonumber\\
  &\qquad + i\left( \oneminboth\bar{\epsilon}_+ \right) \Gamma_0 \Gamma^A \trip{X^A}{\Psi_-}{} + i\bar{\epsilon}_- \Gamma_0 \Gamma^A \trip{X^A}{\Psi_+}{}  \nonumber\\
  \delta_0 A_z(\,\cdot\,) &= i \left( \oneminzbar\bar{\epsilon}_+ \right) \Gamma_z \Gamma_Z \trip{Z}{\Psi_-}{} + i \bar{\epsilon}_- \Gamma_z \Gamma_{\bar{Z}} \trip{\bar{Z}}{\Psi_+}{} \nonumber\\
  &\qquad + i\left( \oneminzbar\bar{\epsilon}_+ \right)\Gamma_z \Gamma^A \trip{X^A}{\Psi_+}{} \nonumber\\
  \delta_0 A_{\bar{z}}(\,\cdot\,) &= i \bar{\epsilon}_- \Gamma_{\bar{z}} \Gamma_Z \trip{Z}{\Psi_+}{} + i \left( \oneminz\bar{\epsilon}_+ \right) \Gamma_{\bar{z}} \Gamma_{\bar{Z}} \trip{\bar{Z}}{\Psi_-}{} \nonumber\\
  &\qquad + i\left( \oneminz\bar{\epsilon}_+ \right)\Gamma_{\bar{z}} \Gamma^A \trip{X^A}{\Psi_+}{}  \nonumber\\[1em]
  \delta_1 Z &= 2i\bar{\epsilon}_- \Gamma_{\bar{Z}}\Psi_- + 2iz \left( \partial \bareps_+ \right) \Gamma_{\bar{Z}}\Psi_+ \\
  \delta_1 \bar{Z} &= 2i\bar{\epsilon}_- \Gamma_{Z}\Psi_- + 2i\bar{z} \left( \bar{\partial} \bareps_+ \right) \Gamma_Z\Psi_+ \nonumber\\
  \delta_1 X^A &= i \left( \left( z\partial + \bar{z}\bar{\partial} \right) \bareps_+ \right) \Gamma^A \Psi_-\nonumber\\
  \delta_1 \Psi_+ &= -\left( D_0 X^A \right) \Gamma_0 \Gamma^A \epsilon_- - \tfrac{1}{6} \trip{X^A}{X^B}{X^C}\Gamma^{ABC}\epsilon_- -\bar{z}\left( D_0 Z \right)\Gamma_0 \Gamma_Z \bar{\partial} \epsilon_+ \nonumber\\
  &\qquad - z\left( D_0 \bar{Z} \right)\Gamma_0 \Gamma_{\bar{Z}} \partial \epsilon_+ + 2z \left( DX^A \right)\Gamma_{\bar{z}}\Gamma^A \partial \epsilon_+ + 2\bar{z} \left( \bar{D}X^A \right)\Gamma_{z}\Gamma^A \bar{\partial} \epsilon_+ \nonumber\\
  &\qquad - \tfrac{1}{2}\bar{z}\trip{X^A}{X^B}{Z}\Gamma^{AB} \Gamma_Z \bar{\partial} \epsilon_+ - \tfrac{1}{2}z\trip{X^A}{X^B}{\bar{Z}}\Gamma^{AB} \Gamma_{\bar{Z}} \partial \epsilon_+ \nonumber\\
  &\qquad - \tfrac{2}{3} X^A \Gamma^A \Gamma_{\bar{z}}\partial \epsilon_+ - \tfrac{2}{3} X^A \Gamma^A \Gamma_{z}\bar{\partial} \epsilon_+ + \tfrac{2}{3} X^A \Gamma^A \Gamma_0 \partial_0 \epsilon_- \nonumber\\
  \delta_1 \Psi_- &= - \left( D_0 X^A \right) \Gamma_0 \Gamma^A \left( z\partial + \bar{z}\bar{\partial}  \right) \epsilon_+ - \tfrac{1}{6} \trip{X^A}{X^B}{X^C} \Gamma^{ABC} \left( z\partial + \bar{z}\bar{\partial} \right)\epsilon_+ \nonumber\\
  \delta_1 A_0(\,\cdot\,) &= i \left( \left( z\partial + \bar{z}\bar{\partial} \right) \bareps_+ \right) \Gamma_0 \Gamma^A \trip{X^A}{\Psi_-}{} \nonumber\\
  \delta_1 A_z(\,\cdot\,) &= i \bar{\epsilon}_- \Gamma_z \Gamma^A \trip{X^A}{\Psi_-}{} + i\bar{z} \left( \bar{\partial} \bareps_+ \right)\Gamma_z \Gamma_Z \trip{Z}{\Psi_-}{} + i\bar{z} \left( \bar{\partial} \bareps_+ \right) \Gamma_z \Gamma^A \trip{X^A}{\Psi_+}{} \nonumber\\
  \delta_1 A_{\bar{z}}(\,\cdot\,) &= i \bar{\epsilon}_- \Gamma_{\bar{z}} \Gamma^A \trip{X^A}{\Psi_-}{} + iz \left( \partial \bareps_+ \right)\Gamma_{\bar{z}} \Gamma_{\bar{Z}} \trip{\bar{Z}}{\Psi_-}{} + iz \left( \partial \bareps_+ \right) \Gamma_{\bar{z}} \Gamma^A \trip{X^A}{\Psi_+}{}\nonumber\ ,
  \label{eq: BLG SUSYs pre-redef}
\end{align}
and $\epsilon=\xi + x^\mu \Gamma_\mu \zeta = \xi + \left( x^0 \Gamma_0 + z\Gamma_z + \bar{z} \Gamma_{\bar{z}} \right)\zeta$.

We see that $S_{-1}$ is not of the form (\ref{eq: simple S_{-1}}), and also that $\delta_{-1}$ acts non-trivially on the bosonic field $A_0$, as well as $\Psi_-$. However, observe that
\begin{align}
  \delta_{-1} A_0 (\,\cdot\,) - \frac{i}{2} \delta_0\trip{Z}{\bar{Z}}{} = 0\ .
\end{align}
Thus, following the discussion in section (\ref{subsec: more general prescription}), we consider the following field redefiniton:
\begin{align}
  A_0 \to \hat{A}_0 = A_0 + \frac{1}{\eta} \frac{i}{2} \trip{Z}{\bar{Z}}{}\ .
  \label{eq: BLG field redef}
\end{align}
We can now calculate how both the action and the supersymmetry field transformations are changed. As described in section \ref{subsec: more general prescription}, we find corrections to $S_{-1}$ arising from terms linear in $\hat{A}_0$ in $S_0$, as well as from terms quadratic in $\hat{A}_0$ in $S_1$. Similarly, $S_0$ is corrected by terms linear in $\hat{A}_0$ in $S_1$.

The resulting shift of the action is
\begin{align}
  S_{-1} &\to S_{-1} + \int d^3 x\, \Big( -\angl{\bar{Z}}{F_{z\bar{z}}(Z)} - \tfrac{1}{8} \angl{\trip{X^A}{Z}{\bar{Z}}}{\trip{X^A}{Z}{\bar{Z}}}
  - \tfrac{1}{4} \angl{\bar{\Psi}_+}{\Gamma_0\trip{Z}{\bar{Z}}{\Psi_+}}\Big)\nonumber\\
  S_0 &\to S_0 + \int d^3 x\, \Big( - \tfrac{i}{2}\angl{D_0 X^A}{\trip{X^A}{Z}{\bar{Z}}} - \tfrac{1}{4}\angl{\bar{\Psi}_-}{\Gamma_0 \trip{Z}{\bar{Z}}{\Psi_-}}\Big)\ ,
\end{align}
while the supersymmetry transformations are shifted by
\begin{align}
  \delta_{-1}\Psi_- &\to \delta_{-1}\Psi_- - \tfrac{i}{2} \trip{X^A}{Z}{\bar{Z}}\Gamma^A \Gamma_0\oneminboth\epsilon_+ \nonumber\\
  \delta_{-1}A_0(\,\cdot\,) &\to \delta_{-1}A_0(\,\cdot\,) +\bar{\epsilon}_+ \Gamma_Z \trip{Z}{\Psi_+}{} - \bar{\epsilon}_+ \Gamma_{\bar{Z}} \trip{\bar{Z}}{\Psi_+}{} \nonumber\\[1em]
  \delta_0\Psi_+ &\to \delta_0\Psi_+ - \tfrac{i}{2}\trip{X^A}{Z}{\bar{Z}}\Gamma^A \Gamma_0\epsilon_- \nonumber\\
  \delta_0\Psi_- &\to \delta_0\Psi_- - \tfrac{i}{2} \trip{X^A}{Z}{\bar{Z}} \Gamma^A \Gamma_0 \left( z\partial + \bar{z}\bar{\partial} \right) \epsilon_+\nonumber\\
  \delta_{0}A_0(\,\cdot\,) &\to \delta_{0}A_0(\,\cdot\,) + \bar{\epsilon}_- \Gamma_Z \trip{Z}{\Psi_-}{} - \bar{\epsilon}_- \Gamma_{\bar{Z}} \trip{\bar{Z}}{\Psi_-}{} \ .  
  \end{align}
In particular we find that 
\begin{align}
  S_{-1} &= \int d^3 x\, \left( -2\angl{\bar{D}Z}{D\bar{Z}}  \right)\ ,
\end{align}
and
\begin{align}
  \delta_{-1} \Psi_- &= 2 \left( \bar{D}Z \right) \Gamma_z \Gamma_Z \oneminzbar\epsilon_+ + 2\left( D\bar{Z} \right)\Gamma_{\bar{z}}\Gamma_{\bar{Z}}\oneminz\epsilon_+\ ,
\end{align}
with $\delta_{-1}$ vanishing on all other fields. 

The theory is now in the correct form to proceed. We introduce Lagrange multiplier field $H$, in the same representation of $\mathcal{G}$ as $Z$, and imposing the condition $\bar{D}Z=0$. The end result is that in the $\eta\to 0$ limit, the theory is described by the action

\begin{align}
 \tilde S &=  \int d^3 x\, \bigg( \frac{1}{2} \angl{D_0 Z}{D_0 \bar{Z}} - 2 \angl{DX^A}{\bar{D}X^A} + \angl{H}{\bar{D}Z} + \angl{\bar{H}}{D\bar{Z}} \nonumber\\
  &\hspace{10mm} - \frac{1}{4} \angl{\trip{X^A}{X^B}{Z}}{\trip{X^A}{X^B}{\bar{Z}}} - \frac{i}{2}\angl{D_0 X^A}{\trip{X^A}{Z}{\bar{Z}}}  -\frac{i}{2} \angl{\bar{\Psi}_+}{\Gamma_0 D_0 \Psi_+} \nonumber\\
  &\hspace{10mm} + 2i \angl{\bar{\Psi}_+}{\left( \Gamma_z \bar{D} + \Gamma_{\bar{z}}D\right)\Psi_-} - \frac{1}{2}\angl{\bar{\Psi}_-}{\Gamma_0 \trip{Z}{\bar{Z}}{\Psi_-}} +i\angl{\bar{\Psi}_+}{\Gamma^A\Gamma_Z\trip{X^A}{Z}{\Psi_-}}  \nonumber\\
  &\hspace{10mm} + i\angl{\bar{\Psi}_+}{\Gamma^A\Gamma_{\bar{Z}}\trip{X^A}{\bar{Z}}{\Psi_-}} + \frac{i}{4} \angl{\bar{\Psi}_+}{\Gamma^{AB}\trip{X^A}{X^B}{\Psi_+}} \nonumber\\
  &\hspace{10mm} + i \big( \left( A_0, F_{z\bar{z}} \right) + \left( A_{\bar{z}}, F_{0z} \right) + \left( A_z, F_{\bar{z}0} \right) + \left( A_0, \left[ A_z, A_{\bar{z}} \right] \right) \big) \bigg)\ ,
  \label{eq: final scaled BLG action}
\end{align}
which is invariant under
\begin{align}
 \tilde \delta Z &= 2i\left( \oneminz\bar{\epsilon}_+ \right) \Gamma_{\bar{Z}}\Psi_+  \nonumber\\
 \tilde \delta \bar{Z} &= 2i\left( \oneminzbar\bar{\epsilon}_+ \right) \Gamma_{Z}\Psi_+  \nonumber\\
 \tilde \delta X^A &= i\left(\oneminboth \bar{\epsilon}_+ \right) \Gamma^A \Psi_- + i\bar{\epsilon}_- \Gamma^A \Psi_+  \nonumber\\
 \tilde \delta \Psi_+ &= - \left( D_0 Z \right) \Gamma_0 \Gamma_Z \oneminzbar\epsilon_+ - \left( D_0 \bar{Z} \right) \Gamma_0 \Gamma_{\bar{Z}} \oneminz\epsilon_+ + 2 \left( DZ \right) \Gamma_{\bar{z}}\Gamma_Z \epsilon_-  \nonumber\\
  &\qquad + 2 \left( \bar{D}\bar{Z} \right) \Gamma_{z}\Gamma_{\bar{Z}} \epsilon_- + 2 \left( DX^A \right) \Gamma_{\bar{z}} \Gamma^A \oneminz\epsilon_+ + 2 \left( \bar{D} X^A \right) \Gamma_{z} \Gamma^A \oneminzbar\epsilon_+  \nonumber\\
  &\qquad -\tfrac{1}{2} \trip{X^A}{X^B}{Z}\Gamma^{AB} \Gamma_Z \oneminzbar\epsilon_+ -\tfrac{1}{2} \trip{X^A}{X^B}{\bar{Z}}\Gamma^{AB} \Gamma_{\bar{Z}} \oneminz\epsilon_+ \nonumber\\
  &\qquad - i\trip{X^A}{Z}{\bar{Z}}\Gamma^A \Gamma_0\epsilon_- + \tfrac{2}{3} Z\Gamma_{\bar{z}} \Gamma_Z \partial\epsilon_- + \tfrac{2}{3} \bar{Z}\Gamma_{z} \Gamma_{\bar{Z}} \bar{\partial}\epsilon_- \nonumber\\
  &\qquad + \tfrac{1}{3}Z\Gamma_Z \Gamma_0 \partial_0 \epsilon_+ + \tfrac{1}{3} \bar{Z}\Gamma_{\bar{Z}}\Gamma_0 \partial_- \epsilon_+ \nonumber\\
 \tilde \delta \Psi_- &= - \left( D_0 X^A \right) \Gamma_0 \Gamma^A \oneminboth\epsilon_+ + 2 \left( D X^A \right) \Gamma_{\bar{z}}\Gamma^A \epsilon_- + 2 \left( \bar{D} X^A \right) \Gamma_{z}\Gamma^A \epsilon_- \nonumber\\
  &\qquad - \left( D_0 Z \right) \Gamma_0\Gamma_Z \epsilon_- -\left( D_0 \bar{Z} \right) \Gamma_0\Gamma_{\bar{Z}} \epsilon_- -\tfrac{1}{6}\trip{X^A}{X^B}{X^C}\Gamma^{ABC}\oneminboth\epsilon_+ \nonumber\\
  &\qquad -\tfrac{1}{2}\trip{X^A}{X^B}{Z}\Gamma^{AB}\Gamma_Z\epsilon_- - \tfrac{1}{2}\trip{X^A}{X^B}{\bar{Z}}\Gamma^{AB}\Gamma_{\bar{Z}}\epsilon_- + 2\bar{z}\left( \bar{D}Z \right) \Gamma_z \Gamma_Z \bar{\partial}\epsilon_+ \nonumber\\
  &\qquad + 2z\left( D\bar{Z} \right) \Gamma_{\bar{z}} \Gamma_{\bar{Z}} \partial\epsilon_+ + \tfrac{2}{3} Z\Gamma_z\Gamma_Z\bar{\partial}\epsilon_+ + \tfrac{2}{3} \bar{Z}\Gamma_{\bar{z}}\Gamma_{\bar{Z}}\partial\epsilon_+ + \tfrac{1}{3} Z\Gamma_Z \Gamma_0 \partial_0 \epsilon_- \nonumber\\
  &\qquad  + \tfrac{1}{3} \bar{Z} \Gamma_{\bar{Z}} \Gamma_0 \partial_0 \epsilon_- - \tfrac{2}{3} X^A \Gamma^A \Gamma_{\bar{z}} \partial \epsilon_- - \tfrac{2}{3} X^A \Gamma^A \Gamma_{z} \bar{\partial} \epsilon_- +\tfrac{1}{3} X^A \Gamma^A \Gamma_0 \partial_0 \epsilon_+ \nonumber\\
  &\qquad - H \Gamma_{\bar{z}} \Gamma_{\bar{Z}}\oneminz \epsilon_+ - \bar{H} \Gamma_z \Gamma_Z \oneminzbar \epsilon_+ \nonumber\\
\tilde  \delta A_0(\,\cdot\,) &= 2\bar{\epsilon}_- \Gamma_Z \trip{Z}{\Psi_-}{} - 2\bar{\epsilon}_- \Gamma_{\bar{Z}} \trip{\bar{Z}}{\Psi_-}{} \nonumber\\
  &\qquad + i\left( \oneminboth\bar{\epsilon}_+ \right) \Gamma_0 \Gamma^A \trip{X^A}{\Psi_-}{} + i\bar{\epsilon}_- \Gamma_0 \Gamma^A \trip{X^A}{\Psi_+}{}  \nonumber\\
\tilde  \delta A_z(\,\cdot\,) &= i \left( \oneminzbar\bar{\epsilon}_+ \right) \Gamma_z \Gamma_Z \trip{Z}{\Psi_-}{} + i \bar{\epsilon}_- \Gamma_z \Gamma_{\bar{Z}} \trip{\bar{Z}}{\Psi_+}{} \nonumber\\
  &\qquad + i\left( \oneminzbar\bar{\epsilon}_+ \right)\Gamma_z \Gamma^A \trip{X^A}{\Psi_+}{} \nonumber\\
 \tilde \delta A_{\bar{z}}(\,\cdot\,) &= i \bar{\epsilon}_- \Gamma_{\bar{z}} \Gamma_Z \trip{Z}{\Psi_+}{} + i \left( \oneminz\bar{\epsilon}_+ \right) \Gamma_{\bar{z}} \Gamma_{\bar{Z}} \trip{\bar{Z}}{\Psi_-}{} \nonumber\\
  &\qquad + i\left( \oneminz\bar{\epsilon}_+ \right)\Gamma_{\bar{z}} \Gamma^A \trip{X^A}{\Psi_+}{}\nonumber\\
\tilde  \delta H &= -2 \left( \oneminzbar\bareps_+ \right)\Gamma_z \Gamma_Z D_0 \Psi_- - i \left( \oneminzbar\bareps_+ \right) \Gamma_z \Gamma_Z \Gamma^{AB} \trip{X^A}{X^B}{\Psi_-}\nonumber\\
  &\qquad -4i \bareps_- \Gamma_Z D\Psi_- -4i \left( \partial \bareps_- \right)\Gamma_Z \Psi_- - 4i\bar{z}\left( \bar{\partial}\bareps_+ \right)\Gamma_Z D\Psi_+ - 2i \bareps_- \Gamma_z \Gamma^A \trip{X^A}{\bar{Z}}{\Psi_-} \nonumber\\
   &\qquad  - 2i \bar{z} \left( \bar{\partial}\bareps_+ \right)\Gamma_z\Gamma_Z \trip{Z}{\bar{Z}}{\Psi_-} -2i\bar{z} \left( \bar{\partial}\bareps_+ \right)\Gamma_z \Gamma^A \trip{X^A}{\bar{Z}}{\Psi_+}\ ,
\end{align}
where
\begin{align}
  \epsilon = \xi + x^\mu \Gamma_\mu \zeta = \xi + \left( x^0 \Gamma_0 + z\Gamma_z + \bar{z} \Gamma_{\bar{z}} \right)\zeta \ .
\end{align}
For $\zeta=0$, we recover the rigid supersymmetry of $S$ as found in \cite{Lambert:2018lgt}. However, in this more general construction, we see that all 32 supersymmetries of the original ${\cal N}=8$ theory survive in the limiting theory. As discussed in \cite{Kucharski:2017jwv} the dynamics is restricted to a three-algebra variation of the Hitchin system for $SU(2)$ gauge group.

\subsection{${\cal N}=6$}\label{subsec: ABJM}

It is a natural question whether we can obtain similar results when applying an analogous scaling to the ABJM/ABJ theory \cite{Aharony:2008ug,Aharony:2008gk}. This is a $U(N)\times U(M)$ Chern-Simons-matter theory with action
\begin{align}
  S = \text{tr} \int & d^3 x \bigg(  - \left( D_\mu Z^I \right) \left( D^\mu Z_I \right) - \frac{8\pi^2}{3k^2}\Upsilon^{KL}_J \Upsilon_{KL}^J\nonumber\\
  &+\frac{k}{4\pi} \epsilon^{\mu\nu\lambda} \left( \left( A_\mu^L \partial_\nu A_\lambda^L - \frac{2i}{3}A_\mu^L A_\nu^L A_\lambda^L \right) - \left( A_\mu^R \partial_\nu A_\lambda^R - \frac{2i}{3}A_\mu^R A_\nu^R A_\lambda^R \right) \right)\nonumber\\
  &-i \bar{\Psi}^I \gamma^\mu D_\mu \Psi_I - \frac{2\pi i}{k}\bar{\Psi}^I [\Psi_I, Z^J;Z_J] + \frac{4\pi i}{k} \bar{\Psi}^I [\Psi_J, Z^J; Z_I]\nonumber\\
  & +\frac{\pi i}{k}\varepsilon_{IJKL} \bar{\Psi}^I [Z^K, Z^L; \Psi^J] - \frac{\pi i}{k}\varepsilon^{IJKL} \bar{\Psi}_I[Z_K, Z_L, \Psi_J] \bigg)\ ,
\end{align}
where we have used the conventions of \cite{Bagger:2008se}. In particular, we have
\begin{align}
  [Z^I, Z^J; Z_K] &= Z^I Z_K Z^J - Z^J Z_K Z^I\nonumber\\
  \Upsilon^{KL}_J &= [Z^K, Z^L; Z_J] - \tfrac{1}{2} \delta_J^K [Z^E, Z^L; Z_E] + \tfrac{1}{2} \delta_J^L [Z^E, Z^K, Z_E]\ .
\end{align}
The matter fields here are $N\times M$ complex matrices, while $A_m^L$ is a hermitian $N\times N$ matrix, and $A_m^R$  a hermitian $M\times M$ matrix. Hermitian conjugation acts to raise/lower the R-symmetry index $I=1,2,3,4$. The fields $Z^I$ and $\psi_I$ transform in the $(\mathbf{N},\bar{\mathbf{M}})$ of the $U(N)_L\times U(M)_R$ gauge symmetry, and so we have
\begin{align}
  D_\mu Z^I = \partial_\mu Z^I - i A_\mu^L Z^I + i Z^I A_\mu^R\ ,
\end{align}
and similarly for $\psi_I$. Finally, we choose conventions in which $\epsilon^{012}=-1$, and a representation for the 3d Clifford algebra is chosen such that the $\gamma^m$, $\mu=0,1,2$ are real $2\times 2$ matrices satisfying $\gamma^0 \gamma^1 \gamma^2 = 1$. 

The theory possesses 12 supercharges corresponding to $\mathcal{N}=6$ rigid supersymmetry, and an additional 12 corresponding to superconformal symmetry. In particular, we find $\delta S=0$ where
\begin{align}
  \delta Z^I &=  i \bar{\epsilon}^{IJ} \psi_J \nonumber\\
\delta A_\mu^L &= \tfrac{2\pi}{k} \left( \bar{\epsilon}^{IJ} \gamma_\mu \Psi_I Z_J - \bar{\epsilon}_{IJ} \gamma_\mu Z^J \Psi^I \right) \nonumber\\
\delta A_\mu^R &= \tfrac{2\pi}{k} \left( \bar{\epsilon}^{IJ} \gamma_\mu Z_J \Psi_I  - \bar{\epsilon}_{IJ} \gamma_\mu \Psi^I Z^J  \right) \nonumber\\
\delta \Psi_J &= D_\mu Z^I \gamma^\mu \epsilon_{IJ} + \tfrac{2\pi}{k} \Upsilon^{KL}_J \epsilon_{KL} + \tfrac{1}{3}Z^J \gamma^\mu \partial_\mu \epsilon_{IJ}\ ,
\end{align}
where the supersymmetry parameter $\epsilon_{IJ}$ takes the form $\epsilon_{IJ}=\xi_{IJ}+x^\mu\gamma_\mu\zeta_{IJ}$ for constant $\xi_{IJ},\zeta_{IJ}$. What's more, $\epsilon_{IJ}$ is anti-symmetric in its R-symmetry indices, and satisfies the reality condition $\epsilon^{IJ}=\frac{1}{2} \varepsilon^{IJKL}\epsilon_{KL}$, which automatically ensures that $\xi_{IJ}$ and $\zeta_{IJ}$ satisfy the same condition. 

Next, we seek a scaling analogous to that performed on the ${\cal N}=8$ theory. In particular, we single out $Z^1$, and look for a scaling which, in the $\eta\to 0$ limit, will localise onto the static $\tfrac{1}{2}$-BPS state $\bar{D} Z^1 = 0$, where we have once again defined the worldvolume complex coordinate $z=x^1+ix^2$, and $D\equiv D_{z}$. Further, we split spinors into definite chirality under $i\gamma^0$, so that $\Psi^\pm_I := P_\pm \Psi_I$ and $\epsilon_{IJ}^\pm := P_\pm \epsilon_{IJ}$. Then, upon comparing the form of $\delta$ to that of the ${\cal N}=8$ theory, we find that $\Psi_1^-$ and $\Psi_A^+$ play a role analogous to that of $\Psi_+$ in the ${\cal N}=8$ theory, while $\Psi_1^+$ and $\Psi_A^-$ are analogous to $\Psi_-$, where here $A=2,3,4$. We find similar correspondences for the constant components of $\epsilon_{IJ}$, giving us the full scalings:
\begin{align}
  x^0 &\to \eta^{-1} x^0 & \xi_{1A}^+ , \xi_{AB}^- &\to \eta^{-1/2} \xi_{1A}^+ , \xi_{AB}^- \nonumber \\
  z,\bar{z} &\to \eta^{-1/2} z,\bar{z} & \xi_{1A}^- , \xi_{AB}^+ &\to \xi_{1A}^- , \xi_{AB}^+\nonumber\\
  Z^1 &\to  Z^1 & \zeta_{1A}^+ , \zeta_{AB}^- &\to \eta^{1/2} \zeta_{1A}^+ , \zeta_{AB}^-\nonumber\\
  Z^A &\to \eta^{1/2} Z^A & \zeta_{1A}^- , \zeta_{AB}^+ &\to \eta \, \zeta_{1A}^- , \zeta_{AB}^+\nonumber\\
  \Psi_1^-,\Psi_A^+ &\to \eta^{1/2} \Psi_1^-,\Psi_A^+ \nonumber \\
  \Psi_1^+,\Psi_A^- &\to \eta \Psi_1^+,\Psi_A^-\ .
  \label{eq: ABJM scaling}
\end{align}
Having performed this scaling, the action takes the form $S=\eta^{-1} S_{-1} + S_0 + \eta S_1$, with
\begin{align}
  S_{-1} &= \text{tr} \int d^3 x \bigg( -2 \left( DZ^1 \right) \left( \bar{D} Z_1 \right) -2 \left( \bar{D}Z^1 \right) \left( D Z_1 \right)  + \tfrac{4\pi^2}{k^2}\trips{Z^1}{Z^A}{Z_1}\trips{Z_1}{Z_A}{Z^1}\nonumber\\
  & \hspace{23mm} - \tfrac{2\pi i}{k} \left( \bar{\Psi}^{A,+} \trips{\Psi_A^+}{Z^1}{Z_1} -\bar{\Psi}^{1,-}\trips{\Psi_1^-}{Z^1}{Z_1} \right) \nonumber\\
  S_0 &=  \text{tr}  \int  d^3 x\, \bigg( \left( D_0 Z^1 \right) \left( D_0 Z_1 \right) - 2 \left( D Z^A \right) \left( \bar{D} Z_A \right) - 2 \left( \bar{D} Z^A \right) \left( D Z_A \right) \nonumber\\
  & +\tfrac{4\pi^2}{3 k^2} \Big( - \trips{Z^A}{Z^1}{Z_A}\trips{Z_B}{Z_1}{Z^B} + 4\trips{Z^A}{Z^1}{Z_B}\trips{Z_A}{Z_1}{Z^B}- \trips{Z^1}{Z^A}{Z_1}\trips{Z_B}{Z_A}{Z^B}\nonumber\\
  &\hspace{14mm} + 2\trips{Z^A}{Z^B}{Z_1}\trips{Z_A}{Z_B}{Z^1} - \trips{Z_1}{Z_A}{Z^1}\trips{Z^B}{Z^A}{Z_B} \Big) \nonumber\\
  &+\tfrac{ki}{2\pi} \Big( \left( A_0^L F_{z\bar{z}}^L + A_{\bar{z}}^L F_{0z}^L + A_z^L F_{\bar{z}0}^L + i A_0^L \left[ A_z^L, A_{\bar{z}}^L \right] \right) - \left( A_0^R F_{z\bar{z}}^R + A_{\bar{z}}^R F_{0z}^R + A_z^R F_{\bar{z}0}^R + i A_0^R \left[ A_z^R, A_{\bar{z}}^R \right] \right) \Big)\nonumber\\
  &+\bar{\Psi}^{1,-} D_0 \Psi_1^- - 2i\bar{\Psi}^{1,+} \gamma_{\bar{z}} D \Psi_1^- - 2i\bar{\Psi}^{1,-} \gamma_{z} \bar{D} \Psi_1^+ -\bar{\Psi}^{A,+} D_0 \Psi_A^+ - 2i\bar{\Psi}^{A,+} \gamma_{\bar{z}} D \Psi_A^- - 2i\bar{\Psi}^{A,-} \gamma_{z} \bar{D} \Psi_A^+\nonumber\\
  &+\tfrac{2\pi i}{k} \Big(  \bar{\Psi}^{1,+} \trips{\Psi_1^+}{Z^1}{Z_1} - \bar{\Psi}^{1,-}\trips{\Psi_1^-}{Z^A}{Z_A} - \bar{\Psi}^{A,-}\trips{\Psi^-_A}{Z^1}{Z_1} \nonumber\\
  &\hspace{14mm} - \bar{\Psi}^{A,+} \trips{\Psi_A^+}{Z^B}{Z_B} + 2\bar{\Psi}^{1,+}\trips{\Psi_A^+}{Z^A}{Z_1} + 2 \bar{\Psi}^{1,-}\trips{\Psi_A^-}{Z^A}{Z_1}\nonumber\\
  &\hspace{14mm} + 2\bar{\Psi}^{A,+} \trips{\Psi_1^+}{Z^1}{Z_A} + 2 \bar{\Psi}^{A,-}\trips{\Psi_1^-}{Z^1}{Z_A} + 2 \bar{\Psi}^{A,+}\trips{\Psi_B^+}{Z^B}{Z_A} \Big)\nonumber\\
  &+\tfrac{4\pi i}{k}\left( \varepsilon_{ABC}\bar{\Psi}^{A,+}\trips{Z^B}{Z^1}{\Psi^{C,-}} - \varepsilon^{ABC} \bar{\Psi}_A^+ \trips{Z_B}{Z_1}{\Psi_C^-} \right) \bigg)\nonumber\\
  S_1 &= \text{tr} \int d^3 x \bigg( \left( D_0 Z^A \right) \left( D_0 Z_A \right) -\bar{\Psi}^{1,+}  D_0 \Psi_1^+ + \bar{\Psi}^{A,-}D_0 \Psi_A^- \nonumber\\
  &\hspace{23mm}+ \tfrac{4\pi^2}{3k^2} \left( 2\trips{Z^B}{Z^C}{Z_A}\trips{Z_B}{Z_C}{Z^A} - \trips{Z^B}{Z^A}{Z_B}\trips{Z_C}{Z_A}{Z^C} \right) \nonumber\\
  & \hspace{23mm}- \tfrac{2\pi i}{k} \left( \bar{\Psi}^{1,+}\trips{\Psi_1^+}{Z^A}{Z_A} + \bar{\Psi}^{A,-}\trips{\Psi_A^-}{Z^B}{Z_B} \right) \bigg)\ .
\end{align}
This satisfies $\delta S=0$, with $\delta = \eta^{-1}\delta_{-1} + \delta_0 + \eta \delta_1$, with
\begin{align}
  \delta_{-1}\Psi_1^+ &= \tfrac{2\pi}{k} \trips{Z^1}{Z^A}{Z_1}\oneminzbar\epsilon_{1A}^+ \\
  \delta_{-1}\Psi_A^- &= 2 \left( \bar{D} Z^1 \right) \gamma_z \oneminzbar \epsilon_{1A}^+ - \tfrac{2\pi}{k} \trips{Z^1}{Z^B}{Z_1} \oneminz \epsilon_{AB}^- \nonumber\\
  \delta_{-1}A_0^L &= \tfrac{2\pi i}{k} \left( - \left( \oneminz \bareps^{1A,+} \right) \Psi_A^+ Z_1 - \left( \oneminzbar \bareps^{+}_{1A} \right) Z^1 \Psi^{A,+} \right) \nonumber\\[1em]
  \delta_{0}Z^1 &= i\left( \oneminz  \bareps^{1A,+} \right)\Psi_A^+  \\
  \delta_{0}Z^A &= -i\left( \oneminz \bareps^{1A,+} \right)\Psi_1^+ - i\bareps^{1A,-}\Psi_1^- + i\bareps^{AB,+} \Psi_B^+ + i\left( \oneminzbar \bareps^{AB,-} \right) \Psi_B^- \nonumber\\
  \delta_{0}\Psi_1^+ &= i\left( D_0 Z^A \right)\oneminzbar\epsilon_{1A}^+ - 2\left( DZ^A \right) \gamma_{\bar{z}} \epsilon_{1A}^- + \tfrac{i}{3}Z^A \partial_0 \epsilon_{1A}^+ -\tfrac{2}{3} Z^A \gamma_{\bar{z}}\partial \epsilon_{1A}^- \nonumber\\
  & \qquad + \tfrac{2\pi}{k} \left( \trips{Z^A}{Z^B}{Z_1}\epsilon_{AB}^+ -\trips{Z^B}{Z^A}{Z_B}\oneminzbar\epsilon_{1A}^+ + \bar{z}\trips{Z^1}{Z^A}{Z_1} \bar{\partial}\epsilon_{1A}^+ \right) \nonumber\\
  \delta_{0}\Psi_1^- &= -2\left( \bar{D}Z^A \right)\gamma_{z}\oneminzbar\epsilon_{1A}^+ +\tfrac{2\pi}{k} \left( \trips{Z^A}{Z^B}{Z_1}\oneminz\epsilon_{AB}^- + \trips{Z^1}{Z^A}{Z_1}\epsilon_{1A}^- \right) \nonumber\\
  \delta_{0}\Psi_A^+ &= -i\left( D_0 Z^1 \right) \oneminzbar \epsilon_{1A}^+ + 2\left( DZ^1 \right) \gamma_{\bar{z}} \epsilon_{1A}^- - 2\left( DZ^B \right) \gamma_{\bar{z}} \oneminz \epsilon_{AB}^- \nonumber\\
  &\qquad +\tfrac{2\pi}{k} \left( \trips{Z^B}{Z^1}{Z_B} \oneminzbar \epsilon_{1A}^+ -2\trips{Z^B}{Z^1}{Z_A}\oneminzbar\epsilon_{1B}^+ - \trips{Z^1}{Z^B}{Z_1}\epsilon_{AB}^+ \right) \nonumber\\
  &\qquad - \tfrac{i}{3} Z^1 \partial_0 \epsilon_{1A}^+ + \tfrac{2}{3} Z^1 \gamma_{\bar{z}}\partial \epsilon_{1A}^-\nonumber\\
  \delta_{0}\Psi_A^- &= i \left( D_0 Z^1 \right) \epsilon_{1A}^- - i \left( D_0 Z^B \right) \oneminz  \epsilon_{AB}^- - 2\left( \bar{D} Z^B \right) \gamma_z\epsilon_{AB}^+  \nonumber\\
  &\qquad + \tfrac{2\pi}{k} \big( \trips{Z^B}{Z^C}{Z_A}\oneminz\epsilon_{BC}^- -2\trips{Z^B}{Z^1}{Z_A}\epsilon_{1B}^- + \trips{Z^B}{Z^1}{Z_B}\epsilon_{1A}^- \nonumber\\
  &\qquad - \trips{Z^1}{Z^B}{Z_1} \oneminz\epsilon_{AB}^- -z\trips{Z^1}{Z^B}{Z_1} \partial\epsilon_{AB}^- \big)  \nonumber\\
  &\qquad +\tfrac{i}{3} Z^1 \partial_0 \epsilon_{1A}^- - \tfrac{i}{3} Z^B \partial_0 \epsilon_{AB}^- - \tfrac{2}{3}Z^B \gamma_z \bar{\partial} \epsilon_{AB}^+ + 2\bar{z} \left( \bar{D}Z^1 \right) \gamma_z \bar{\partial}\epsilon_{1A}^+ + \tfrac{2}{3} Z^1 \gamma_z \bar{\partial}\epsilon_{1A}^+ \nonumber\\
  \delta_{0}A_0^L &= \tfrac{2\pi i}{k} \big( \left( \oneminz\bareps^{1A,+} \right) \Psi_1^+ Z_A - \bareps^{1A,-} \Psi_1^- Z_A + \bareps^{1A,-} \Psi_A^- Z_1 + \bareps^{AB,+} \Psi_A^+ Z_B \nonumber\\
  &\qquad - \left( \oneminzbar\bareps^{AB,-} \right) \Psi_A^- Z_B + \left( \oneminzbar \bareps^{+}_{1A} \right) Z^A \Psi^{1,+} - \bareps^{-}_{1A} Z^A \Psi^{1,-} + \bareps^{-}_{1A} Z^1 \Psi^{A,-} \nonumber\\
  &\qquad + \bareps^{+}_{AB} Z^B\Psi^{A,+} - \left( \oneminz\bareps^{-}_{AB} \right) Z^B\Psi^{A,-} -z \left( \partial \bareps^{1A,+} \right) \Psi_A^+ Z_1 - \bar{z} \left( \bar{\partial}\bareps_{1A}^+ \right) Z^1 \Psi^{A,+} \big) \nonumber\\
  \delta_{0}A_z^L &= \tfrac{2\pi}{k} \left( -\bareps^{1A,-}\gamma_z \Psi_A^+ Z_1 + \left( \oneminzbar\bareps^{AB,-} \right)\gamma_z \Psi_A^+ Z_B + \left( \oneminzbar\bareps_{1A}^+ \right) \gamma_z \left( Z^1 \Psi^{A,-} - Z^A \Psi^{1,-} \right) \right) \nonumber\\
  \delta_{0}A_{\bar{z}}^L &= \tfrac{2\pi}{k} \left( \bareps_{1A}^- \gamma_{\bar{z}} Z^1 \Psi^{A,+} - \left( \oneminz\bareps_{AB}^- \right) \gamma_{\bar{z}}Z^B \Psi^{A,+} - \left( \oneminz \bareps^{1A,+} \right)\gamma_{\bar{z}}\left( \Psi_A^- Z_1 - \Psi_1^- Z_A \right)  \right) \nonumber\\[1em]
  \delta_{1} Z^1 &= i \bareps^{1A,-} \Psi_A^- + iz \left( \partial \bareps^{1A,+} \right) \Psi_A^+ \\
  \delta_{1} Z^A &= -iz \left( \partial \bareps^{1A,+} \right)\Psi_1^+ + i \bar{z} \left( \bar{\partial}\bareps^{AB,-} \right)\Psi_B^- \nonumber\\
  \delta_{1} \Psi_1^+ &= i\bar{z} \left( D_0 Z^A \right) \bar{\partial}\epsilon_{1A}^+ - \tfrac{2\pi}{k} \bar{z} \trips{Z^B}{Z^A}{Z_B}\bar{\partial}\epsilon_{1A}^+ \nonumber\\
  \delta_{1} \Psi_1^- &= - i \left( D_0 Z^A \right) \epsilon_{1A}^- + \tfrac{2\pi}{k} \left( - \trips{Z^B}{Z^A}{Z_B}\epsilon_{1A}^- + z \trips{Z^A}{Z^B}{Z_1} \partial \epsilon_{AB}^- \right) - \tfrac{i}{3} Z^A \partial_0 \epsilon_{1A}^- \nonumber\\
  &\qquad -2\bar{z} \left( \bar{D}Z^A \right)\gamma_z \bar{\partial}\epsilon_{1A}^+ - \tfrac{2}{3}Z^A \gamma_z \bar{\partial}\epsilon_{1A}^+ \nonumber\\
  \delta_{1} \Psi_A^+ &= i \left( D_0 Z^B \right) \epsilon_{AB}^+ + \tfrac{2\pi}{k} \left( \trips{Z^B}{Z^C}{Z_A}\epsilon_{BC}^+ + \bar{z} \trips{Z^B}{Z^1}{Z_B}\bar{\partial}\epsilon_{1A}^+ - 2 z \trips{Z^B}{Z^1}{Z_A}\bar{\partial}\epsilon_{1B}^+ \right) \nonumber\\
  &\qquad + \tfrac{i}{3} Z^B \partial_0 \epsilon_{AB}^+ -i\bar{z} \left( D_0 Z^1 \right) \bar{\partial}\epsilon_{1A}^+ - \tfrac{2}{3} Z^B \gamma_{\bar{z}} \partial \epsilon_{AB}^- \nonumber\\
  \delta_{1} \Psi_A^- &= \tfrac{2\pi}{k} z \left( \trips{Z^B}{Z^C}{Z_A}\partial \epsilon_{BC}^- - \trips{Z^1}{Z^B}{Z_1} \partial \epsilon_{AB}^- \right) - iz \left( D_0 Z^B \right) \partial \epsilon_{AB}^- \nonumber\\
  \delta_{1}A_0^L &= \tfrac{2\pi i}{k} \left( z \left( \partial \bareps^{1A,+}\right) \Psi_1^+ Z_A - \bar{z} \left( \bar{\partial}\bareps^{AB,-} \right) \Psi_A^- Z_B + \bar{z} \left( \bar{\partial} \bareps_{1A}^+ \right) Z^A \Psi^{1,+} - z \left( \partial \bareps_{AB}^- \right) Z^B \Psi^{A,-} \right) \nonumber\\
  \delta_{1}A_z^L &= \tfrac{2\pi}{k} \big( \bareps^{1A,-} \gamma_z \Psi_1^+ Z_A - \bareps_{AB}^+ \gamma_z Z^B \Psi^{A,-} + \bar{z} \left( \bar{\partial} \bareps^{AB,-} \right) \gamma_z \Psi_A^+ Z_B \nonumber\\
  &\qquad + \bar{z} \left( \bar{\partial}\bareps_{1A}^+ \right) \gamma_z\left( Z^1 \Psi^{A,-} -  Z^A \Psi^{1,-} \right) \big) \nonumber\\
  \delta_{1}A_{\bar{z}}^L &= \tfrac{2\pi}{k} \big( -\bareps_{1A}^- \gamma_{\bar{z}} Z^A \Psi^{1,+} + \bareps^{AB,+} \gamma_{\bar{z}} \Psi^{-}_A Z_B  - z \left( \partial \bareps^{-}_{AB} \right) \gamma_{\bar{z}} Z^B \Psi^{A,+} \nonumber\\
  &\qquad - z \left( \partial\bareps^{1A,+} \right) \gamma_{\bar{z}} \left( \Psi^{-}_A  Z_1 - \Psi^{-}_1 Z_A \right) \big)\ ,
\end{align}
where one obtains $\delta A_m^R$ by swapping the order of all fields in the expressions for $\delta A_m^L$. Here, $\epsilon_{IJ}$ is once again of the form
\begin{align}
  \epsilon_{IJ} = \xi_{IJ} + x^\mu\gamma_\mu \zeta_{IJ} = \xi_{IJ} + \left( x^0\gamma_0 + z\gamma_z + \bar{z} \gamma_{\bar{z}} \right) \zeta_{IJ}\ .
\end{align}
As in the ${\cal N}=8$ case, we see that neither $S_{-1}$ or $\delta_{-1}$ are in the simple form required to directly find the limiting theory. However, we note that
\begin{align}
  \delta_{-1}A_0^L + \frac{2\pi}{k} \delta_0\left( Z^1 Z_1 \right) = 0 , \qquad 
  \delta_{-1}A_0^R + \frac{2\pi}{k} \delta_0\left( Z_1 Z^1 \right) = 0 \ ,
\end{align}
Thus, we consider the field redefiniton
\begin{align}
  A_0^L \to \hat{A}_0^L = A_0^L - \frac{1}{\eta} \frac{2\pi}{k} Z^1 Z_1 ,\qquad 
  A_0^R \to \hat{A}_0^R = A_0^R - \frac{1}{\eta} \frac{2\pi}{k} Z_1 Z^1 \ .
\end{align}
The shift to the action is then
\begin{align}
  S_{-1} &\to S_{-1} + \text{tr} \int d^3 x\,\, \Big( -2i Z^1 Z_1 F_{z\bar{z}}^L + 2i Z_1 Z^1 F_{z\bar{z}}^R - \tfrac{4\pi^2}{k^2}\trips{Z^1}{Z^A}{Z_1}\trips{Z_1}{Z_A}{Z^1} \nonumber\\
  & \hspace{38mm}+ \tfrac{2\pi i}{k} \left( \bar{\Psi}^{A,+} \trips{\Psi_A^+}{Z^1}{Z_1} -\bar{\Psi}^{1,-}\trips{\Psi_1^-}{Z^1}{Z_1} \right) \Big) \nonumber\\
  S_0 &\to S_0 + \text{tr} \int d^3 x\,\, \Big( -\tfrac{2\pi i}{k}\left( \left( D_0 Z_A \right) \trips{Z^A}{Z^1}{Z_1} + \left( D_0 Z^A \right) \trips{Z_A}{Z_1}{Z^1} \right) \nonumber\\
  & \hspace{38mm} +\tfrac{2\pi i}{k}  \left( \bar{\Psi}^{1,+} \trips{\Psi_1^+}{Z^1}{Z_1}  - \bar{\Psi}^{A,-}\trips{\Psi^-_A}{Z^1}{Z_1} \right) \Big)\ ,
\end{align}
while the supersymmetry transformations are shifted by
\begin{align}
  \delta_{-1} \Psi_1^+ &\to \delta_{-1}\Psi_1^+ - \tfrac{2\pi}{k} \trips{Z^1}{Z^A}{Z_1}\oneminzbar \epsilon_{1A}^+ \nonumber  \\
  \delta_{-1} \Psi_A^- &\to \delta_{-1} \Psi_A^- + \tfrac{2\pi}{k} \trips{Z^1}{Z^B}{Z_1}\oneminz \epsilon_{AB}^-\nonumber\\
  \delta_{-1} A_0^L &\to A_0^L + \tfrac{2\pi i}{k} \left( \left( \oneminz \bareps^{1A,+} \right) \Psi_A^+ Z_1 + \left( \oneminzbar \bareps^{+}_{1A} \right) Z^1 \Psi^{A,+} \right)\nonumber\\[1em]
  \delta_0 \Psi_1^+ &\to \delta_0 \Psi_1^+ - \tfrac{2\pi}{k}\bar{z}\trips{Z^1}{Z^A}{Z_1} \bar{\partial}\epsilon_{1A}^+ \nonumber \\
  \delta_0 \Psi_1^- &\to \delta_0 \Psi_1^- + \tfrac{2\pi}{k}\trips{Z^1}{Z^A}{Z_1} \epsilon_{1A}^- \nonumber\\
  \delta_0 \Psi_A^+ &\to \delta_0 \Psi_A^+ - \tfrac{2\pi}{k}\trips{Z^1}{Z^B}{Z_1} \epsilon_{AB}^+ \nonumber\\
  \delta_0 \Psi_A^- &\to \delta_0 \Psi_A^- + \tfrac{2\pi}{k}z\trips{Z^1}{Z^B}{Z_1} \partial\epsilon_{AB}^- \nonumber\\
  \delta_0 A_0^L &\to \delta_0 A_0^L + \tfrac{2\pi i}{k} \left( \bareps^{1A,-}\Psi_A^- Z_1 + \bareps_{1A}^- Z^1 \Psi^{A,-} + z \left( \partial \bareps^{1A,+} \right) \Psi_A^+ Z_1 + \bar{z} \left( \bar{\partial}\bareps_{1A}^+ \right) Z^1 \Psi^{A,+} \right)\ ,
\end{align}
where again the shift to $\delta A_0^R$ can be determined from that of $\delta A_0^L$ by simply swapping all pairs of fields. 
In particular, we now have that
\begin{align}
  S_{-1} = \text{tr} \int d^3 x\,\, \Big( -4 \left( \bar{D} Z^1 \right) \left( D Z_1 \right) \Big), \qquad 
    \delta_{-1}\Psi_A^- &= 2 \left( \bar{D} Z^1 \right) \gamma_z \oneminzbar \epsilon_{1A}^+\ ,
\end{align}
with $\delta_{-1}=0$ on all other fields. This is now in the correct form to proceed. The end result is that in the $\eta\to 0$ limit, the theory is described by the action

\begin{align}
   \tilde S &=  \text{tr}  \int  d^3 x\, \bigg( \left( D_0 Z^1 \right) \left( D_0 Z_1 \right) - 2 \left( D Z^A \right) \left( \bar{D} Z_A \right) - 2 \left( \bar{D} Z^A \right) \left( D Z_A \right) + H \left( DZ_1 \right) + \bar{H}\left( \bar{D}Z^1 \right) \nonumber\\
  & -\tfrac{2\pi i}{k}\left( \left( D_0 Z_A \right) \trips{Z^A}{Z^1}{Z_1} + \left( D_0 Z^A \right) \trips{Z_A}{Z_1}{Z^1} \right) \nonumber\\
  & +\tfrac{4\pi^2}{3 k^2} \Big( - \trips{Z^A}{Z^1}{Z_A}\trips{Z_B}{Z_1}{Z^B} + 4\trips{Z^A}{Z^1}{Z_B}\trips{Z_A}{Z_1}{Z^B}- \trips{Z^1}{Z^A}{Z_1}\trips{Z_B}{Z_A}{Z^B}\nonumber\\
  &\hspace{14mm} + 2\trips{Z^A}{Z^B}{Z_1}\trips{Z_A}{Z_B}{Z^1} - \trips{Z_1}{Z_A}{Z^1}\trips{Z^B}{Z^A}{Z_B} \Big) \nonumber\\
  &+\tfrac{ki}{2\pi} \Big( \left( A_0^L F_{z\bar{z}}^L + A_{\bar{z}}^L F_{0z}^L + A_z^L F_{\bar{z}0}^L + i A_0^L \left[ A_z^L, A_{\bar{z}}^L \right] \right) - \left( A_0^R F_{z\bar{z}}^R + A_{\bar{z}}^R F_{0z}^R + A_z^R F_{\bar{z}0}^R + i A_0^R \left[ A_z^R, A_{\bar{z}}^R \right] \right) \Big)\nonumber\\
  &+\bar{\Psi}^{1,-} D_0 \Psi_1^- - 2i\bar{\Psi}^{1,+} \gamma_{\bar{z}} D \Psi_1^- - 2i\bar{\Psi}^{1,-} \gamma_{z} \bar{D} \Psi_1^+ -\bar{\Psi}^{A,+} D_0 \Psi_A^+ - 2i\bar{\Psi}^{A,+} \gamma_{\bar{z}} D \Psi_A^- - 2i\bar{\Psi}^{A,-} \gamma_{z} \bar{D} \Psi_A^+\nonumber\\
  &+\tfrac{2\pi i}{k} \Big(  \bar{\Psi}^{1,+} \trips{\Psi_1^+}{Z^1}{Z_1} - \bar{\Psi}^{1,-}\trips{\Psi_1^-}{Z^A}{Z_A} - \bar{\Psi}^{A,-}\trips{\Psi^-_A}{Z^1}{Z_1} \nonumber\\
  &\hspace{14mm} - \bar{\Psi}^{A,+} \trips{\Psi_A^+}{Z^B}{Z_B} + 2\bar{\Psi}^{1,+}\trips{\Psi_A^+}{Z^A}{Z_1} + 2 \bar{\Psi}^{1,-}\trips{\Psi_A^-}{Z^A}{Z_1}\nonumber\\
  &\hspace{14mm} + 2\bar{\Psi}^{A,+} \trips{\Psi_1^+}{Z^1}{Z_A} + 2 \bar{\Psi}^{A,-}\trips{\Psi_1^-}{Z^1}{Z_A} + 2 \bar{\Psi}^{A,+}\trips{\Psi_B^+}{Z^B}{Z_A} \Big)\nonumber\\
  &+\tfrac{4\pi}{k}\left( \varepsilon_{ABC}\bar{\Psi}^{A,+}\trips{Z^B}{Z^1}{\Psi^{C,-}} - \varepsilon^{ABC} \bar{\Psi}_A^+ \trips{Z_B}{Z_1}{\Psi_C^-} \right) \bigg)\ ,
\end{align}
which preserves the full 24 supersymmetries of the original ${\cal N}=6$ theory. In particular, we have $\tilde \delta \tilde S=0$, with
\begin{align}
    \tilde\delta Z^1 &= i\left( \oneminz  \bareps^{1A,+} \right)\Psi_A^+ \nonumber\\
 \tilde \delta Z^A &= -i\left( \oneminz \bareps^{1A,+} \right)\Psi_1^+ - i\bareps^{1A,-}\Psi_1^- + i\bareps^{AB,+} \Psi_B^+ + i\left( \oneminzbar \bareps^{AB,-} \right) \Psi_B^- \nonumber\\
  \tilde\delta \Psi_1^+ &= i\left( D_0 Z^A \right)\oneminzbar\epsilon_{1A}^+ - 2\left( DZ^A \right) \gamma_{\bar{z}} \epsilon_{1A}^- + \tfrac{i}{3}Z^A \partial_0 \epsilon_{1A}^+ -\tfrac{2}{3} Z^A \gamma_{\bar{z}}\partial \epsilon_{1A}^- \nonumber\\
  & \qquad + \tfrac{2\pi}{k} \left( \trips{Z^A}{Z^B}{Z_1}\epsilon_{AB}^+ -\trips{Z^B}{Z^A}{Z_B}\oneminzbar\epsilon_{1A}^+  \right) \nonumber\\
 \tilde \delta \Psi_1^- &= -2\left( \bar{D}Z^A \right)\gamma_{z}\oneminzbar\epsilon_{1A}^+ +\tfrac{2\pi}{k} \left( \trips{Z^A}{Z^B}{Z_1}\oneminz\epsilon_{AB}^- + 2\trips{Z^1}{Z^A}{Z_1}\epsilon_{1A}^- \right) \nonumber\\
  \tilde\delta \Psi_A^+ &= -i\left( D_0 Z^1 \right) \oneminzbar \epsilon_{1A}^+ + 2\left( DZ^1 \right) \gamma_{\bar{z}} \epsilon_{1A}^- - 2\left( DZ^B \right) \gamma_{\bar{z}} \oneminz \epsilon_{AB}^- \nonumber\\
  &\qquad +\tfrac{2\pi}{k} \left( \trips{Z^B}{Z^1}{Z_B} \oneminzbar \epsilon_{1A}^+ -2\trips{Z^B}{Z^1}{Z_A}\oneminzbar\epsilon_{1B}^+ - 2\trips{Z^1}{Z^B}{Z_1}\epsilon_{AB}^+ \right) \nonumber\\
  &\qquad - \tfrac{i}{3} Z^1 \partial_0 \epsilon_{1A}^+ + \tfrac{2}{3} Z^1 \gamma_{\bar{z}}\partial \epsilon_{1A}^-\nonumber\\
  \tilde\delta \Psi_A^- &= i \left( D_0 Z^1 \right) \epsilon_{1A}^- - i \left( D_0 Z^B \right) \oneminz  \epsilon_{AB}^- - 2\left( \bar{D} Z^B \right) \gamma_z\epsilon_{AB}^+  + \tfrac{2\pi}{k} \big( \trips{Z^B}{Z^C}{Z_A}\oneminz\epsilon_{BC}^- \nonumber\\
  &\qquad -2\trips{Z^B}{Z^1}{Z_A}\epsilon_{1B}^- + \trips{Z^B}{Z^1}{Z_B}\epsilon_{1A}^- - \trips{Z^1}{Z^B}{Z_1} \oneminz\epsilon_{AB}^-  \big)  +\tfrac{i}{3} Z^1 \partial_0 \epsilon_{1A}^- \nonumber\\
  &\qquad - \tfrac{i}{3} Z^B \partial_0 \epsilon_{AB}^- - \tfrac{2}{3}Z^B \gamma_z \bar{\partial} \epsilon_{AB}^+ + 2\bar{z} \left( \bar{D}Z^1 \right) \gamma_z \bar{\partial}\epsilon_{1A}^+ + \tfrac{2}{3} Z^1 \gamma_z \bar{\partial}\epsilon_{1A}^+ -\tfrac{1}{2}H \gamma_z \oneminzbar \epsilon_{1A}^+ \nonumber\\
 \tilde \delta A_0^L &= \tfrac{2\pi i}{k} \big( \left( \oneminz\bareps^{1A,+} \right) \Psi_1^+ Z_A - \bareps^{1A,-} \Psi_1^- Z_A + 2\bareps^{1A,-} \Psi_A^- Z_1 + \bareps^{AB,+} \Psi_A^+ Z_B \nonumber\\
  &\qquad - \left( \oneminzbar\bareps^{AB,-} \right) \Psi_A^- Z_B + \left( \oneminzbar \bareps^{+}_{1A} \right) Z^A \Psi^{1,+} - \bareps^{-}_{1A} Z^A \Psi^{1,-} + 2\bareps^{-}_{1A} Z^1 \Psi^{A,-} \nonumber\\
  &\qquad + \bareps^{+}_{AB} Z^B\Psi^{A,+} - \left( \oneminz\bareps^{-}_{AB} \right) Z^B\Psi^{A,-} \big) \nonumber\\
 \tilde \delta A_z^L &= \tfrac{2\pi}{k} \left( -\bareps^{1A,-}\gamma_z \Psi_A^+ Z_1 + \left( \oneminzbar\bareps^{AB,-} \right)\gamma_z \Psi_A^+ Z_B + \left( \oneminzbar\bareps_{1A}^+ \right) \gamma_z \left( Z^1 \Psi^{A,-} - Z^A \Psi^{1,-} \right) \right) \nonumber\\
 \tilde \delta A_{\bar{z}}^L &= \tfrac{2\pi}{k} \left( \bareps_{1A}^- \gamma_{\bar{z}} Z^1 \Psi^{A,+} - \left( \oneminz\bareps_{AB}^- \right) \gamma_{\bar{z}}Z^B \Psi^{A,+} - \left( \oneminz \bareps^{1A,+} \right)\gamma_{\bar{z}}\left( \Psi_A^- Z_1 - \Psi_1^- Z_A \right)  \right) \nonumber\\
 \tilde \delta H &= -2 \left( \oneminz \bareps^{1A,+}  \right)\gamma_{\bar{z}} \left( D_0 \Psi_A^- - \tfrac{2\pi i}{k}\trips{\Psi_A^-}{Z^B}{Z_B} \right) - 4i \left( \partial \bareps^{1A,-} \right)\Psi_A^- - 4i \bareps^{1A,-}\bar{D}\Psi_A^- \nonumber\\
  &\qquad -4i z \left( \partial \bareps^{1A,+} \right) \bar{D} \Psi_A^+ + \tfrac{8\pi i}{k} \big( \bareps^{AB,+}\gamma_{\bar{z}}\trips{\Psi_A^-}{Z^1}{Z_B} - \bareps_{1A}^- \gamma_{\bar{z}}\trips{Z^A}{Z^1}{\Psi^{1,+}} \nonumber\\
  & \qquad -z \left( \partial \bareps_{AB}^- \right) \gamma_{\bar{z}} \trips{Z^B}{Z^1}{\Psi^{A,+}} -z \left( \partial \bareps^{1A,+} \right)\gamma_{\bar{z}}\left( \trips{\Psi_A^-}{Z^1}{Z_1}-\trips{\Psi_1^-}{Z^1}{Z_A} \right)\big)\ ,
\end{align}
where once again, $\tilde\delta A_\mu^R$ is obtained from $\tilde\delta A_\mu^L$ by swapping the order of all pairs of fields, and $\epsilon_{IJ}$ is given by
\begin{align}
  \epsilon_{IJ} = \xi_{IJ} + x^\mu \gamma_\mu \zeta_{IJ} = \xi_{IJ} +\left( x^0\gamma_0 + z\gamma_z + \bar{z} \gamma_{\bar{z}} \right)\zeta_{IJ}\ ,
\end{align}
with $\xi_{IJ}$, $\zeta_{IJ}$, and hence $\epsilon_{IJ}$ satisfying the reality condition $\epsilon^{IJ}=\tfrac{1}{2} \varepsilon^{IJKL}\epsilon_{KL}$. We have introduced the $N\times M$ complex matrix $H$ transforming in the $\left( \mathbf{N}, \bar{\mathbf{M}} \right)$ of $U(N)_L\times U(M)_R$, which acts a a Lagrange multiplier imposing $\bar{D}Z^1=0$. Its Hermitian conjugate, transforming in the $\left(\bar{\mathbf{N}}, \mathbf{M} \right)$, is denoted $\bar{H}$.

As with the ${\cal N}=8$ case above  there are two constraints. The first comes from integrating out $\bar H$ which simply implies
\begin{align}
\bar D Z^1 = 0\ .
\end{align}
However there is also the Gauss law constraint that comes from integrating out $A^{L/R}_0$:
\begin{align}
F^L_{z\bar z}    &= \frac{ 2\pi^2 i}{k^2}(Z^A[Z_A,Z_1;Z^1]- [Z^A,Z^1;Z_1]Z_A)\nonumber\\
 F^R_{z\bar z}  &= -\frac{ 2\pi^2 i}{k^2}([Z_A,Z_1;Z^1]Z^A- Z_A[Z^A,Z^1;Z_1])\ ,
\end{align}
where for simplicity  we have set the fermions to zero and assumed static configurations with $D_0Z^1=0$. These are a bi-fundamental version of the Hitchin equation.

\section{Conclusion}

In this paper we   presented a general procedure in which we induce a non-Lorentzian rescaling of a field theory and take the limit where the scale parameter vanishes. The resulting action has a divergent term but we showed that it can be removed and replaced with a Lagrange multiplier term  in such a way that supersymmetry is preserved. The new action also has a Lifshitz scaling symmetry.  The dynamics are then restricted to the zeros of the divergent term so that the Manton approximation of slow motion on a moduli space becomes exact, in a manner reminiscent of localization calculations.   We provided explicit examples of this procedure: maximally supersymmetric Yang-Mills, leading to a five-dimensional theory with 16 supersymmetries and 8 superconformal symmetries, and  maximally supersymmetric Chern-Simons-matter theory leading to a three-dimensional theory with 32 super(conformal) symmetries, as first constructed in  \cite{Lambert:2018lgt}. We also  extended the analysis to   ABJM/ABJ models with 24 super(conformal) symmetries. 

It would be interesting to consider further examples such as theories with less supersymmetry but more varied matter content. In addition there appears to be a natural interpretation of this construction within an AdS/CFT context for M5-branes \cite{Lambert:2019jwi}. We hope to report on these issues in due course.

\section*{Acknowledgements}

N.L. was in part supported by the STFC grant ST/P000258/1. R.M. is supported an the STFC studentship ST10837.

\section*{Appendix: Scaling from the point of view of the M5-brane}

In this appendix, we motivate the scaling of five-dimensional maximally supersymmetric Yang-Mills as described in section \ref{sec: M5s} from the perspective of the M5-brane. In particular, we argue that this scaling arises naturally by considering the action on multiple coincident M5-branes in a Lorentz-boosted frame that is almost null, with boost parameter $\beta\approx 1-\tfrac{1}{4}\eta^2$. As discussed in \cite{Lambert:2011gb}, the limiting theory obtained in section \ref{sec: M5s} reduces to quantum mechanics on instanton moduli space. Thus in this limit we reproduce the DLCQ description of M5-branes \cite{Aharony:1997an,Aharony:1997th}.\\

The set-up is as follows (see also \cite{Bilal:1999ff}). Consider $(10+1)$ Minkowski spacetime  with coordinates $\{x^M\}$, $M=0,1,\dots,10$. Let $x^5$ be the coordinate on the M-Theory circle, with
\begin{align}
  x^5\equiv x^5+2\pi\eta R\ ,
  \label{eq: x^5 ident}
\end{align}
for a dimensionless parameter $\eta$. Suppose there are a stack of $N$ coincident M5-branes spanning $\{x^0,\dots,x^5\}$. In the limit that $\eta$ is small, these branes are described by the action on $N$ coincident D4-branes spanning $\{x^0,\dots,x^4\}$ - namely five-dimensional maximally supersymmetric Yang-Mills. Letting $\Sigma$ be the submanifold  of spacetime defined by $x^5, x^6, \dots,x^{10}=0$, we have the action
\begin{align}
  S= \frac{1}{g^2} \text{tr} \int_\Sigma d^5 x\,\, \bigg( -\frac{1}{4} F_{\mu\nu}F^{\mu\nu} - \frac{1}{2} \left( D_\mu X^I \right)\left( D^\mu X^I \right) + \frac{i}{2} \bar{\Psi}\Gamma^\mu D_\mu \Psi \nonumber\\
  -\frac{1}{2} \bar{\Psi} \Gamma_5 \Gamma^I \left[ X^I,\Psi \right] + \frac{1}{4} \left[ X^I, X^J \right] \left[ X^I, X^J \right] \bigg)\ ,
\end{align}
where $\mu=0,\dots,4$ and $I=6,\dots,10$. Here, the $X^I$ and their superpartners $\Psi$ are $N\times N$ real matrices. The spinors are 32-component spinors of the ambient eleven-dimensional spacetime, satisfying $\Gamma_{012345}\Psi = -\Psi$. Finally, we have $D_\mu X^I = \partial_\mu X^I - i\left[ A_\mu, X^I \right]$, and similarly for $\Psi$. The coupling $g$ is related to the M-theory radius by
\begin{align}
  g^2 = 4\pi^2 \eta R = 2\pi\times \left( 2\pi \eta R \right)\ .
\end{align}
This action is invariant under the supersymmetries (\ref{eq: YM SUSYs}). We now choose a different set of coordinates for the eleven-dimensional spacetime, and determine the form of $S$ with respect to them. In particular, consider the Lorentz-equivalent choice of coordinates given by
\begin{align}
  \begin{pmatrix}
  	\tilde{x}^0\\
  	\tilde{x}^5
  \end{pmatrix} =
  \frac{1}{\sqrt{1-\beta^2}}
  \begin{pmatrix}
  	x^0 - \beta x^5\\
  	x^5 - \beta x^0
  \end{pmatrix},\qquad
  \beta = \frac{1-\tfrac{1}{2}\eta^2}{1+\tfrac{1}{2}\eta^2}\ ,
\end{align}
with $\tilde{x}^M=x^M$ for $M\neq 0,5$. The usual lightcone coordinates in this new basis are  
\begin{align}
  \tilde{x}^+ &= \tfrac{1}{\sqrt{2}}\left( \tilde{x}^0 + \tilde{x}^5 \right) = \tfrac{\eta}{2}\left( x^0 + x^5 \right)\nonumber\\
  \tilde{x}^- &= \tfrac{1}{\sqrt{2}}\left( \tilde{x}^0 - \tilde{x}^5 \right) = \tfrac{1}{\eta}\left( x^0 - x^5 \right)\ ,
\end{align}
and hence the identification (\ref{eq: x^5 ident}) appears as
\begin{align}
  \begin{pmatrix}
  	\tilde{x}^+\\
  	\tilde{x}^-
  \end{pmatrix} \equiv 
  \begin{pmatrix}
  	\tilde{x}^+\\
  	\tilde{x}^-
  \end{pmatrix} + 
  \begin{pmatrix}
  	\pi\eta^2 R\\
  	-2\pi R
  \end{pmatrix}\ .
\end{align}
Finally, we make the further shift to the coordinates, given by
\begin{align}
  \hat{x}^+ &= \tilde{x}^+ +\tfrac{1}{2}\eta^2 \tilde{x}^- \quad\left( = \eta x^0 \right)\nonumber\\
  \hat{x}^- &=\tilde{x}^-\ ,
\end{align}
again with $\hat{x}^M = x^M$ for all $M\neq 0,5$. Thus, $\hat{x}^-$ has period $2\pi R$, while $\hat{x}^+$ is non-compact. In this basis, the eleven-dimensional Minkowski metric   is $ds^2 = -d\hat{x}^+ d\hat{x}^- + \eta^2 \left( d\hat{x}^- \right)^2 + d\hat{x}^i d\hat{x}^i + d\hat{x}^I d\hat{x}^I$, and so in the $\eta\to 0$ limit $\hat{x}^\pm$ are true lightcone coordinates.

Finally, we write $S$ in terms of these new coordinates $\{\hat{x}^M\}$. This is a non-trivial calculation, since the $\hat{x}^M$ are not Lorentz-equivalent to our original $x^M$. In particular, we write $S$ in a generally covariant form in terms of the pullbacks of tensor fields defined in a neighbourhood of $\Sigma$, while the inclusion of spinors also requires the introduction of a frame bundle $e^M_\mu$.   We also rescale the scalars $X^I\to \eta X^I$, along with the same rescaling for their superpartners $\Psi$. From a purely field-theoretic perspective this can be seen as bringing the spatial kinetic terms for the $X^I$ to canonical normalisation, while geometrically we are really rescaling the coordinates transverse to the brane as $x^I\to \eta x^I$. Indeed, it is easy to show that this shift can be induced by such a coordinate shift in the ambient spacetime. We also expand out any $\Gamma$-matrices in terms of constant matrices in the tangent space, as $\Gamma_\mu=e_\mu^M \Gamma_M$.

Dropping the hats on coordinates, the end result is the action
\begin{align}
  S=\eta^{-1} S_{-1} + S_0 + \eta S_1\ ,
\end{align}
with
\begin{align}
  S_{-1} &= \frac{1}{g^2} \text{tr}\int d^4 x\, dx^+ \left( -\frac{1}{4} F_{ij} F_{ij} \right)  \nonumber\\ 
  S_0 &=  \frac{1}{g^2} \text{tr}\int d^4 x\, dx^+ \Bigg( \frac{1}{2} F_{+i} F_{+i} - \frac{1}{2} \left( D_i X^I \right) \left( D_i X^I \right) - \frac{i}{2} \bar{\Psi} \Gamma_- D_+ \Psi \nonumber\\
  &\hspace{36mm}  +\frac{i}{2} \bar{\Psi} \Gamma_i D_i \Psi + \frac{1}{2} \bar{\Psi} \Gamma_- \Gamma^I [X^I,\Psi] \Bigg)\nonumber\\
  S_{1} &= \frac{1}{g^2} \text{tr}\int d^4 x\, dx^+ \Bigg( \frac{1}{2} \left( D_+ X^I \right) \left( D_+ X^I \right) - \frac{i}{4}\bar{\Psi} \Gamma_+ D_0 \Psi \nonumber\\
  &\hspace{36mm} -\frac{1}{4} \bar{\Psi} \Gamma_+ \Gamma^I [X^I, \Psi] + \frac{1}{4} [X^I, X^J][X^I, X^J] \Bigg)\ .
\end{align}
with supersymmetries as in (\ref{eq: YM scaled SUSYs}), with $x^0\to x^+$. This is precisely the action (\ref{eq: YM scaled}) we found by naively scaling five-dimensional maximally supersymmetric Yang-Mills, except now the $x^+$ coordinate in the ambient spacetime plays the role of time on the worldvolume. In particular, in the $\eta\to 0$ limit, this coordinate becomes a regular lightcone coordinate on spacetime.

%\bibliography{../../Refs.bib}
%\bibliographystyle{JHEP}
%\bibliographystyle{plain}

\end{document}